\documentclass[prb,showpacs,preprintnumbers,amsfonts,amssymb,amsmath,floats,twocolumn,aps,superscriptaddress]{revtex4-1}

\usepackage{amsmath}
\usepackage{amssymb}
\usepackage{graphicx}
\usepackage{bm}
\usepackage{bbold}
\usepackage{xcolor}
\usepackage[caption=false]{subfig}
\usepackage[pdftex]{hyperref} %
\definecolor{darkblue}{rgb}{0,0,.5} %
\definecolor{black}{rgb}{0,0,0} %
\hypersetup{
  pdftex=true,
  colorlinks=true, breaklinks=true, linkcolor=darkblue,
  menucolor=darkblue,
  urlcolor=darkblue, citecolor=darkblue }

\def\mc{\mathcal}
\def\bs{\boldsymbol}

\def\Im{\operatorname{Im}}

\def\op{}
\def\mat{\bs}
\def\fcl{\widehat}

\def\cre#1#2{#1^\dagger_{#2}}%
\def\ann#1#2{#1^{\vphantom{\dagger}}_{#2}}%

\def\cc#1{\cre{c}{#1}}%
\def\ac#1{\ann{c}{#1}}%
\def\cf#1{\cre{f}{#1}}%
\def\af#1{\ann{f}{#1}}%
\def\nf#1{n^f_{#1}}%

\def\ra{\rightarrow}%
\def\ua{\uparrow}%
\def\da{\downarrow}%

\def\com[#1,#2]{\left[#1,#2\right]}
\def\contcom[#1,#2]{\left[#1\stackrel{\circ}{,}#2\right]}

\def\ev<#1>{\langle#1\rangle}

\def\bra<#1|{\left<#1\right|}
\def\ket|#1>{\left|#1\right>}
\def\braket<#1|#2|#3>{\langle#1|#2|#3\rangle}

\begin{document}
  
\title{Time-dependent Mott transition in the periodic Anderson model with nonlocal hybridization}

\author{Felix Hofmann} \email{fhofmann@physik.uni-hamburg.de}
\author{Michael Potthoff} 
\affiliation{I. Institut f\"ur Theoretische Physik, Universit\"at Hamburg, Jungiusstra\ss{}e 9, 20355 Hamburg, Germany}

\begin{abstract}
The time-dependent Mott transition in a periodic Anderson model with off-site, nearest-neighbor hybridization is studied within the framework of nonequilibrium self-energy functional theory. 
Using the two-site dynamical-impurity approximation, we compute the real-time dynamics of the optimal variational parameter and of different observables initiated by sudden quenches of the Hubbard-$U$ and identify the critical interaction. 
The time-dependent transition is orbital selective, i.e., in the final state, reached in the long-time limit after the quench to the critical interaction, the Mott gap opens in the spectral function of the localized orbitals only. 
We discuss the dependence of the critical interaction and of the final-state effective temperature on the hybridization strength and point out
the various similarities between the nonequilibrium and the equilibrium Mott transition.
It is shown that these can also be smoothly connected to each other by increasing the duration of a $U$-ramp from a sudden quench to a quasi-static process.
The physics found for the model with off-site hybridization is compared with the dynamical Mott transition in the single-orbital Hubbard model and with the dynamical crossover found for the real-time dynamics of the conventional Anderson lattice with on-site hybridization.
\end{abstract}



\maketitle

\section{Introduction}
\label{sec:introduction}

Lattice models of strongly correlated electrons in the thermodynamical
limit typically exhibit a rich phenomenology with transitions between
several equilibrium phases controlled by different model parameters or
temperature.  The correlation-driven metal-insulator (Mott)
transition\citep{mott1949,mott1990,gebhard2010} in the single-band
Hubbard model at half filling represents a prototypical example which
has been studied extensively by means of dynamical mean-field theory
(DMFT).\citep{georges1996,kotliar2004} Recently, there has been
renewed theoretical interest in the Mott transition in the context of
real-time dynamics following a quantum
quench.\citep{eckstein2009b,schiro2010b,hamerla2013,hamerla2014} Such
quench dynamics in the fermionic Hubbard model is, at least in
principle, in reach of experiments done with ultra-cold atomic gases
trapped in optical
lattices.\citep{esslinger2010,bloch2012,lewenstein2012}

Generically, isolated quantum systems driven out of equilibrium are
believed to show
thermalization,\citep{srednicki1994,deutsch1991,rigol2008} i.e., in
the long-time limit time-averaged values of all relevant observables
are given by an average within a thermal ensemble. Thermalization is
known to be delayed if the Hamiltonian is close to an integrable point
in the parameter space. In this case the system does not thermalize
directly but gets temporarily trapped in an intermediate prethermal
state.\citep{berges2004,moeckel2008,moeckel2010,kollar2011,marcuzzi2013}

Using nonequilibrium dynamical mean-field theory
(NE-DMFT),\citep{schmidt2002,freericks2006,aoki2013} prethermalization
has been found and discussed for the half-filled Hubbard
model.\citep{eckstein2009b,eckstein2010b} After suddenly switching on
the Hubbard interaction from $U=0$ to $U=U_{\rm fin}$,
prethermalization plateaus develop for weak $U_{\rm fin}$ and damped
collapse-and-revival oscillations for strong $U_{\rm fin}$.  Only for
final interactions close to a characteristic ``critical'' interaction
$U^{\rm dyn}_{\rm c}$, a rapid thermalization of the double occupancy
and of the jump of the momentum-distribution function is found. This
``dynamical Mott transition'' has also been studied within various
other approaches.\citep{schiro2010b,hamerla2013,hamerla2014,hofmann2016}

Within the dynamical mean-field approach, it is presently not clear if
and how the ``transition'' at $U^{\rm dyn}_{\rm c}$ is related to the
conventional (equilibrium) Mott transition at $U = U_{\rm c}(T)$ and
temperature $T$.  On the one hand, the equilibrium state that is
reached after the quench to $U^{\rm dyn}_{\rm c}$ is characterized by
a temperature $T_{\rm eff}$ which is more than an order of magnitude
higher than the critical temperature $T_{\rm c}$ for the equilibrium
Mott transition above which there is merely a smooth metal-insulator
crossover in the equilibrium phase
diagram.\citep{eckstein2009b,eckstein2010b} This implies that the
parametric distance in the equilibrium phase diagram between the
thermal state reached after the quench, i.e., at ($U^{\rm dyn}_{\rm
  c}, T_{\rm eff}$), and the critical point $(U_{\rm c}(T_{\rm c}),
T_{\rm c})$ is large, and hence that there is no obvious interrelation
between the two phenomena.

On the other hand, a link is suggested by a study using the
time-dependent Gutzwiller variational method \citep{sandri2012} where
the instantaneous $U$-quench has been replaced by a ramp with a
characteristic time $\Delta t_{\rm ramp}$. It has been found that
there is a well-defined $U^{\rm dyn}_{\rm c}$ for any $\Delta t_{\rm
  ramp}$, and that in the limit $\Delta t_{\rm ramp} \to \infty$,
where the ramp can be considered as a quasi-stationary thermodynamical
process, the ``critical'' interaction $U^{\rm dyn}_{\rm c}$ approaches
$U_{\rm c} \equiv U_{\rm c}(T=0)$.  Exactly the same phenomenology
could be verified recently\citep{hofmann2016} using the two-site
dynamical impurity approximation (DIA) which is constructed within the
framework of the nonequilibrium generalization \citep{hofmann2013}
of self-energy functional theory (SFT). \cite{potthoff2003,potthoff2003c}

The purpose of the present paper is to employ the nonequilibrium
two-site DIA to study the same question for a two-orbital model where
the critical interaction for the Mott transition is expected to depend
on additional parameters and where a more systematic study of the
interrelation between the equilibrium Mott transition and the
``dynamical'' transition can be carried out.  There are only a few
NE-DMFT studies beyond a single-orbital model.
\citep{werner2012,aoki2013,werner2015} None of them, however, is
addressing the Mott transition.  Most probably this is due to the fact
that an efficient solver for the time-dependent effective impurity
problem within the NE-DMFT is not yet available if, like for the Mott
transition, perturbative approaches should be disregarded.
Continuous-time quantum Monte-Carlo techniques\citep{gull2011}
represent a notable exception.  Unfortunately, the dynamical sign (or
phase) problem severely restricts the accessible propagation time.
Exact-diagonalization-based and related
techniques\citep{gramsch2013,balzer2015,wolf2014} have not yet been
applied to the two-orbital case, and cluster-perturbation
methods\citep{potthoff2011c,gramsch2015} are still lacking internal
consistency and self-consistent feedback necessary to address phase
transitions within an advanced mean-field framework.  The
time-dependent slave-boson mean-field approach represents a promising
alternative.  For the two-orbital Hubbard model, \citep{behrmann2013}
however, the Mott-Hubbard phenomenology turns out to be much more
complicated as compared to the single-orbital case, and a systematic
study of the dynamical Mott transition and its parameter dependencies
has not yet been performed.

In the present paper, we will discuss the results of a study of a
variant of the periodic Anderson model (PAM). While in the PAM with
on-site hybridization there is a smooth crossover from a hybridization
band insulator at $U=0$ to a strongly correlated Kondo insulator for
strong $U$, it has been suggested by \citet{huscroft1999} and verified
in different studies \citep{held2000,held2000b,vandongen2001} using
DMFT, linearized DMFT,\citep{bulla2000} and the Gutzwiller
approximation\citep{brinkman1970} that there is a quantum-critical
point $U_{\rm c}$ at $T=0$ in a model variant with nearest-neighbor
hybridization.  This is caused by the fact that the Kondo effect is
absent for weak hybridization strengths since in reciprocal space the
momentum-dependent hybridization exactly vanishes at the Fermi surface
of the half-filled noninteracting model.\citep{vandongen2001} In this
model, the conduction-electron system stays metallic while the
localized-electron system undergoes a Mott transition at $U_{\rm c}$.
Close to the transition, the evolution of the corresponding spectral
function is very similar to that of the single-orbital Hubbard
model.\citep{held2000b} In particular, the quasi-particle weight $Z
\to 0$ for $U \to U_{\rm c}$ opposed to the model with on-site
hybridization which does not show a transition and where $Z \to 0$
only for $U\to \infty$.  Furthermore, for the nearest-neighbor case,
the critical interaction strongly depends on the hybridization
strength $V$ with $U_{\rm c} \approx \mbox{const.}\times V^2$. For our
study, we consider the PAM in the paramagnetic state at half filling
for both, on-site and nearest-neighbor hybridization.

The paper is organized as follows: In Section~\ref{sec:model-method}
we introduce the model variants and briefly discuss the two-site DIA.
This is used in Sec.~\ref{sec:equilibrium} to study the equilibrium phase diagram.  
In Sec.~\ref{sec:nonequilibrium}, the nonequilibrium two-site DIA is employed to 
address the dynamical Mott transition in the real-time dynamics following an interaction quench or ramp.
The conclusions are summarized in Sec.~\ref{sec:conclusions}.

\section{Model and method}
\label{sec:model-method}

We consider the half-filled periodic Anderson model.  
The Hamiltonian is given by
\begin{multline}
  \label{eq:PAMHamiltonian}
  \op H(t) = -T_{\rm hop} \sum_{\langle ij
    \rangle,\sigma}\cc{i\sigma}\ac{j\sigma} + \sum_{ij \sigma} V_{ij}
  (\cc{i\sigma}\af{j\sigma} + \mathrm{h.c.}) \\ + U(t) \sum_{i}
  \left(\nf{i\ua}-\frac{1}{2}\right)\left(\nf{i\da}-\frac{1}{2}\right)
  \, .
\end{multline}
Here, $c^{(\dagger)}_{i\sigma}$ annihilates (creates) a conduction
electron at site $i$ with spin projection $\sigma=\uparrow,
\downarrow$.  Conduction electrons hop with amplitude $-T_{\rm hop}$ between
neighboring sites $\langle ij \rangle$ of a lattice.  To fix the
energy and time scale we set $T_{\rm hop} \equiv 1$.  Further,
$f^{(\dagger)}_{i\sigma}$ is the annihilator (creator) of an electron
in a localized $f$ orbital, $\nf{i\sigma} = \cf{i\sigma}\af{i\sigma}$
is the corresponding occupation-number operator, and $U(t)$ is the
time-dependent strength of the local Hubbard interaction on the $f$
orbitals.  We will consider interaction quenches and ramps to induce
nontrivial real-time dynamics.  The $c$ and $f$ subsystems are coupled
via a hybridization term.  Two cases are studied: (a) orbitals are
coupled via a hybridization between nearest neighbors, i.e., $V_{ij} =
V$ for neighboring sites $i$ and $j$ and zero otherwise and (b)
on-site hybridization $V_{ij} = V\delta_{ij}$.

\begin{figure}[b]
  \centering%
  \def\figheight{1.26cm}%
  \subfloat[nearest-neighbor
  hybridization\label{fig:PAMoffsite}]{{\includegraphics[height=\figheight]{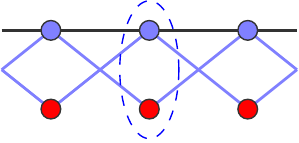}}}
  \hfill \subfloat[on-site
  hybridization\label{fig:PAMonsite}]{{\includegraphics[height=\figheight]{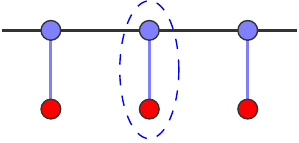}}}%
  \hfill \subfloat[reference
  system\label{fig:PAMref}]{{\includegraphics[height=\figheight]{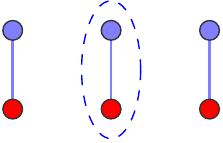}}}%
  \hfill
  \caption{Illustration of the PAM with (a) nearest-neighbor and (b)
    on-site hybridization.  Blue filled dots: uncorrelated sites
    ($c$ orbitals), red filled dots: correlated sites with $U>0$
    ($f$ orbitals).  Black lines: nearest-neighbor hopping between
    $c$ orbitals.  Blue lines: hybridization between $c$ and
    $f$ orbitals.  Both cases (a) and (b) are treated within the
    two-site DIA; the building block of the respective reference
    system, as shown in (c), is highlighted by a dotted blue ellipse.}
  \label{fig:PAMs}
\end{figure}

As it is numerically more convenient, the model is studied on the one-dimensional lattice, see Fig.~\ref{fig:PAMs}.  
One should note, however, that because of the mean-field approach used here, the lattice dimension is not really relevant for our study.  
For a bipartite lattice, numerical results will in first place depend on the variance of the noninteracting density of states, and the lattice dimension mainly enters via the coordination number only.
Clearly, the mean-field approach is best justified for high-dimensional lattices.

We compute equilibrium and time-dependent properties by means of the
dynamical impurity approximation (DIA) within the self-energy
functional theory (SFT): \cite{hofmann2013,hofmann2015} The
self-energy is approximated by the exact self-energy $\mat\Sigma'$ of
a reference system with a small Hilbert space.  The reference system
must share the same interaction part with the original system but its
one-particle parameters are variationally optimized.  With the
two-site DIA we consider the simplest meaningful reference system
which has been proven to qualitatively capture the essence of the
dynamical Mott transition in the single-orbital Hubbard
model. \cite{hofmann2016} The reference system consists of a set of
decoupled two-site systems.  Each building block is given by a single
correlated ($f$) site and a single uncorrelated ($c$) bath site
coupled via a time-dependent hybridization strength $V'(t)$, see
Fig.~\ref{fig:PAMs}.  At half filling the reference-system Hamiltonian reads
\begin{multline}
  \label{eq:refPAMHamiltonian}
  \op H'(t) = V'(t) \sum_{i\sigma} (\cc{i\sigma}\af{i\sigma} +
  \mathrm{h.c.}) \\ + U(t)
  \sum_i\left(\nf{i\ua}-\frac{1}{2}\right)\left(\nf{i\da}-\frac{1}{2}\right)
  \, .
\end{multline}
Here and in the following, primed quantities refer to the reference
system.  The trial self-energy generated by the Hamiltonian
(\ref{eq:refPAMHamiltonian}) is \emph{local}, and the resulting
two-site DIA is a single-site mean-field approximation, similar to the
nonequilibrium DMFT but with a much simpler bath.  The numerical
computation of $\mat\Sigma'$ can therefore be performed with
conceptually simple exact-diagonalization techniques.

In the extreme case of vanishing hopping, the reference system would provide the exact self-energy of the original model with on-site hybridization, Fig.~\ref{fig:PAMs}(b), since the reference and the original system are identical for $T_{\rm hop}=0$.  
Contrary, the model with nearest-neighbor hybridization remains nontrivial for $T_{\rm hop}=0$ (it decouples into two chains with alternating correlated and uncorrelated sites).
It is therefore tempting to expect that the two-site DIA is more adequate in the parameter regime $T_{\rm hop} \ll V, U$.
The present study, however, is done for hybridization strengths $V \sim T_{\rm hop}$.
Hence, there is no {\em a priori} reason to assume that the reference system generates a more reliable approximation for the on-site case. 

Optimization of time-dependent parameters relies on the fact that the
grand potential of the original system $\Omega$ can generally be
expressed as a functional $\fcl\Omega[\mat\Sigma]$ of the
nonequilibrium self-energy which is stationary at the \emph{physical}
self-energy $\mat\Sigma$, i.e., $\delta \fcl\Omega[\mat\Sigma] =
0$. \cite{hofmann2013} Here, $\mat\Sigma$ is defined on the
Keldysh-Matsubara contour $\mc C$,\citep{leeuwen2006c,rammer2007}
i.e., has elements $\Sigma_{ij,\sigma}(z,z')$ with complex contour
times $z,z'$.  The essential point is that $\fcl\Omega[\mat\Sigma]$
can be evaluated exactly by numerical means on the subspace of trial
self-energies $\mat\Sigma'$, generated by the reference system and
thus parametrized by $V'$.  At each instant of time the optimal
hybridization strength $V'_{\rm opt}(t)$ can be determined according
to
\begin{equation}
  \label{eq:EulerEqn}
  \left.\frac{\delta\fcl\Omega[\mat\Sigma']}{\delta
      V'(z)}\right|_{V'(t) = V_{\rm opt}'(t)} = 0\,.
\end{equation}
Note that the on-site energies of the two-site reference system,
opposed to the hybridization, are entirely fixed by the manifest
particle-hole symmetry at half filling.

Formally, variations have to be carried out independently on the upper
and lower Keldysh branch, i.e., variations must be carried out with
respect to $V'(z)$ where $z$ is the complex contour time.  In
practice, we make use of the inherently causal structure of the Euler
equation (\ref{eq:EulerEqn}) which allows us to set up a
time-propagation scheme for $V'_{\rm opt}(t)$.  It is beneficial to
analytically calculate the functional derivative in
Eq.~(\ref{eq:EulerEqn}) and to numerically solve the resulting
root-finding problem.
All time-dependent observables are finally derived from the
approximate DIA Green's function $\mat G^{\rm DIA} = (\mat G_0^{-1} -
\mat\Sigma'_{\rm opt})^{-1}$, where $\mat G_0$ is the free Green's
function of the original system and $\mat\Sigma'_{\rm opt}$ is the
self-energy of the reference system evaluated at $V'_{\rm opt}$.
Details of the general SFT framework and of its numerical
implementation can be found in
Refs.~\onlinecite{hofmann2013,hofmann2015}.

Calculations for the PAM with nearest-neighbor and with on-site
hybridization have been performed for a one-dimensional lattice of
$L=40$ sites and with periodic boundary conditions.  This is
sufficient for convergence of the results presented here as has been
routinely checked by performing calculations for different $L$.  The
inverse temperature $\beta$ sets the length of the Matsubara branch in
all contour integrations.  Contour integrals are computed using
higher-order schemes with imaginary time steps $\Delta\tau \leq 0.2$.
For the real-time propagation up to $t_{\rm max} \lesssim 25$ (in units of the inverse hopping) we obtain converged results only for significantly shorter time steps $\Delta t \leq 0.04$ since,
opposed to the equilibrium case, we are limited to the trapezoidal
rule for all integrations along the Keldysh branches (cf.\
Refs.~\onlinecite{hofmann2013,hofmann2015} for details).

\section{Equilibrium}
\label{sec:equilibrium}

\begin{figure}[t]
  \centering%
  \includegraphics[width=\columnwidth]{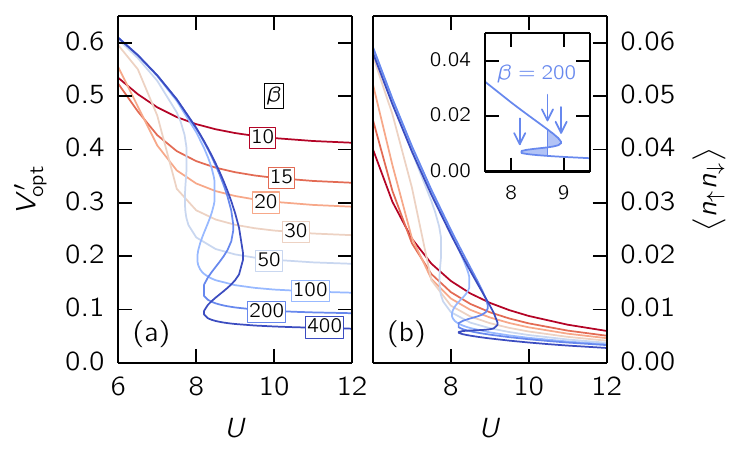}
  \caption{Equilibrium results for the PAM with nearest-neighbor
    hybridization $V \approx 0.866$ ($V^2 = 0.75$): (a) $U$-dependence
    of the optimal variational parameter for different inverse temperatures
    $\beta$ as indicated and (b) corresponding $f$ double
    occupancies. The inset in (b) shows the Maxwell construction at
    $\beta = 200$: the mid arrow indicates the critical interaction
    $U_{\rm c}$, the outer arrows point at the spinodal points, which
    define the region where metallic and insulating solutions
    coexist.}
  \label{fig:offsiteVD}
\end{figure}

We first turn our attention to the equilibrium properties of the PAM
with nearest-neighbor hybridization [see
Fig.~\subref*{fig:PAMoffsite}].  For small inverse temperature $\beta$
and weak $U$ the (time-independent) optimal variational parameter $V'_{\rm opt}$
is easily found by a global search and can then be traced through the
entire $\beta$--$U$ parameter space by a local search for which we use
Broyden's method. \citep{kelley1987,broyden1965}

Fig.~\ref{fig:offsiteVD}(a) shows results for $V'_{\rm opt}$ as
function of $U$ for fixed hybridization $V \approx 0.866$ ($V^2 =
0.75$) in the original system and different $\beta$.  At a
sufficiently low temperatures (high $\beta$), three coexisting
solutions for $V'_{\rm opt}$ are found for certain values of $U$.
This indicates a phase transition of first order (see discussion
below).

To better understand the details of the transition, we extrapolate the
results for $V'_{\rm opt}$ to zero temperature: For weak $U$ the
optimal variational parameter becomes temperature independent and converges to a
finite value, whereas for strong $U$, carrying out the $T\ra 0$ limit
we find $V'_{\rm opt} \ra 0$. Thus, there is a quantum-critical point at
$T=0$ at an intermediate interaction $U_{\rm c}$ for which the bath site in the
reference system decouples, i.e., for $U \to U_{\rm c}$ the
zero-temperature optimal hybridization $V'_{\rm opt} \to 0$.
  
\begin{figure}[t]
  \centering%
  \includegraphics[width=\columnwidth]{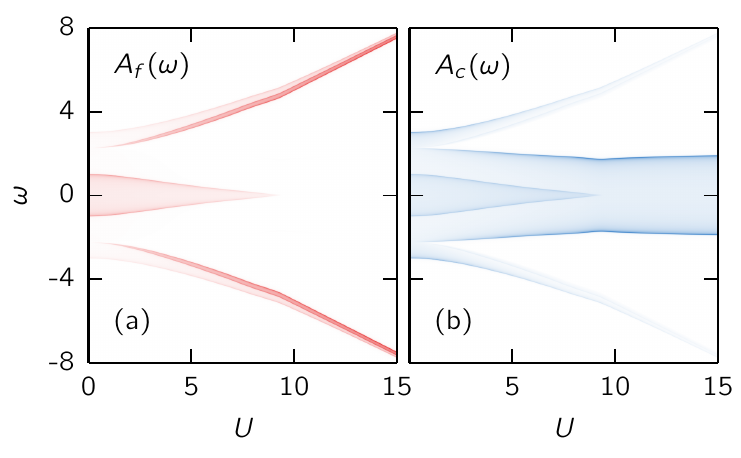}
  \caption{Local spectral densities on the $f$ and $c$ orbitals at
    zero temperature and nearest-neighbor hybridization $V^{2}=0.75$
    as functions of $U$.  }
  \label{fig:offsiteA}
\end{figure}

The physical meaning of the critical point can be uncovered by
computing the zero-temperature one-particle spectral function.  To
this end, we first exploit the translational symmetry of the lattice
and use Fourier transformation to calculate the one-particle
dispersion in reciprocal space:
\begin{equation}
  {\boldsymbol \varepsilon}(k) =  \begin{pmatrix}
    \varepsilon_c(k) & V(k) \\ V(k) & 0
  \end{pmatrix} \, .
\end{equation}
Here, $\varepsilon_c(k) = - 2 T \cos(ka)$ and $V(k) = 2 V \cos(ka)$
(the lattice constant $a$ is set to unity).  Note that the
momentum-dependent hybridization exactly vanishes at the Fermi surface
of the half-filled noninteracting model for the original system with
nearest-neighbor hybridization [Fig.~\subref*{fig:PAMoffsite}].
Opposed to this case, we have $V(k) = V$ for the system with on-site
hybridization [Fig.~\subref*{fig:PAMonsite}].  The $f$ self-energy of
the two-site reference system is given by \citep{lange1998}
\begin{equation}
  \label{eq:SE2srefsys}
  \Sigma'(\omega) = \frac{U^2}{4}\frac{\omega}{\omega^2 - 9{V'}^2} \,.
\end{equation}
Evaluating this at the optimal variational parameter $V'_{\rm opt}$, the DIA
Green's function is obtained from
\begin{align}
  \label{eq:Gsftk}
  \mat G^{\rm DIA}(k,\omega) &=
  \begin{pmatrix} \omega - \varepsilon_c(k) & -V(k) \\
    -V(k) & \omega - \Sigma'_{\rm opt}(\omega)\end{pmatrix}^{-1} \,.
\end{align}
Finally, the local spectral density is derived via the general
relation $\mat A(\omega) = -\frac{1}{\pi}\lim_{\eta\ra 0}\Im \mat
G(\omega+i\eta)$, from the $k$-summed Green's function $\mat G(\omega) = L^{-1} \sum_k
\mat G(k,\omega)$.  For the numerical calculation we have used a
finite but small value $\eta = 0.01$ which slightly broadens the
$\delta$-peaks of the spectral density.  Results for the local,
orbital-resolved $f$ and $c$ spectral densities $A_f(\omega) \equiv
A_{ff}(\omega)$ and $A_c(\omega) \equiv A_{cc}(\omega)$ are shown in
Fig.~\ref{fig:offsiteA}.

In the noninteracting case both, the $f$ and the $c$ spectral density are finite at $\omega=0$, and the system is a normal metal.
The spectral densities have the same support but strongly different weights.
Note that, opposed to the case of the model with on-site hybridization, there is no hybridization band gap opening at $\omega = 0$ since the $k$-dependent hybridization vanishes at the Fermi wave vectors $k=\pm \pi / 2$.  
For finite $U$, a three-peak structure develops in $A_f(\omega)$ with two Hubbard-like peaks located at $\omega\approx\pm U/2$ and a central quasi-particle resonance at $\omega = 0$.

In the strong-coupling limit $U\to \infty$, we have $V'_{\rm opt} = 0$
and thus the self-energy has a single pole at $\omega=0$ with a strong
weight $U^{2}/4$.  This Hubbard-I-type self-energy leads to a
Mott-insulating $f$-electron spectral density with a large gap of the
order of $U$ while the $c$-electron system becomes dynamically
decoupled from the $f$-electron system and the $c$ spectral density
approaches the form of the noninteracting $c$ spectral density for
$U\to \infty$, i.e., the $c$-electron system remains metallic.

\begin{figure}[t]
  \centering%
  \includegraphics[width=\columnwidth]{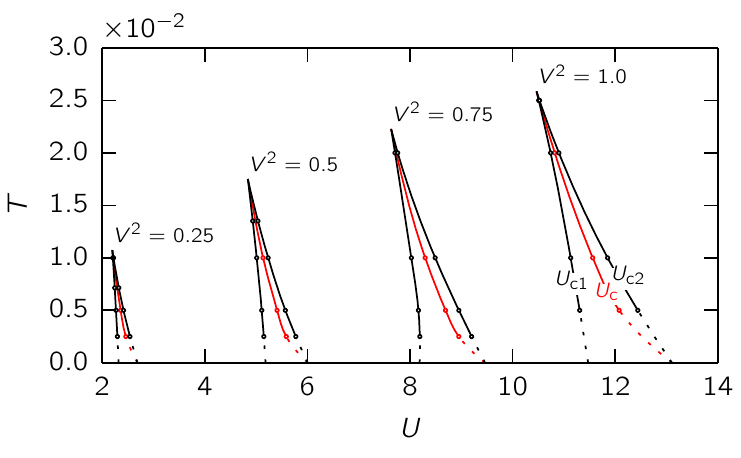}
  \caption{Orbital-selective Mott transition in the $T$-$U$ phase
    diagram of the half-filled PAM for different nearest-neighbor
    hybridizations $V$ as indicated.  Below some critical temperature,
    there is a coexistence of a metallic phase with an
    orbital-selective Mott insulator in a $U$ range between $U_{c1}$
    and $U_{c2}$ (black lines). Red line: first-order phase boundary
    $U_c(T)$.  Extrapolations of the results to zero temperature are
    indicated by dashed lines.}
  \label{fig:pds}
\end{figure}

In fact, there is an orbital-selective Mott metal-insulator transition
at intermediate coupling strength $U_{\rm c} \approx 9.46$. For $U\to
U_{\rm c}$ the optimal hybridization $V'_{\rm opt} \to 0$, i.e., the
bath site in the reference system decouples.  This implies that the
two poles of the self-energy at $\omega = \pm 3 V'_{\rm opt}$ merge
and that the quasi-particle resonance in the $f$-electron spectrum
vanishes in a pre-formed Mott-Hubbard gap (see
Fig.~\ref{fig:offsiteA}(a)).  This is the typical scenario of the Mott
transition in the single-band Hubbard model as obtained within the
full DMFT \cite{georges1996} or within the two-site
DIA. \cite{potthoff2003b,hofmann2016} On the other hand, the
$c$-electron spectral function stays gapless when $U \to U_{\rm c}$
(see Fig.~\ref{fig:offsiteA}(b)).  For $U > U_{\rm c}$, the system is
characterized by an orbital-selective Mott phase with localized $f$
electrons and itinerant $c$ electrons.  This fully agrees with the
findings of Refs.\ \onlinecite{held2000,held2000b,vandongen2001}.

We have calculated the full $T$-$U$ phase diagram of the
orbital-selective Mott transition for different hybridization
strengths, see Fig.~\ref{fig:pds}.  Below a certain critical
temperature $T_{\rm c}$, we find three coexisting solutions in a
certain range of interactions $U_{c1}(T) < U < U_{c2}(T)$.  The
respective $f$ double occupancy $d = \ev<\nf{\ua}\nf{\da}>$ is
depicted in Fig.~\ref{fig:offsiteVD}(b).  
From the hysteresis behavior of $d(U)$ we can infer the critical interaction $U_{\rm c}(T)$ of the
phase transition by a Maxwell construction, as shown in the inset of
Fig.~\ref{fig:offsiteVD}(b).  $U_{\rm c}(T)$ is located within the
coexistence region.  The phase transition at $U_{\rm c}(T)$ is first
order for any temperature $0 < T < T_{\rm c}$.  At $T=0$ the
transition is continuous.  At $T=T_{\rm c}$ the first-order line
terminates in a second-order critical end point above which the
transition is a smooth crossover.

\begin{figure}[t]
  \centering%
  \includegraphics[width=\columnwidth]{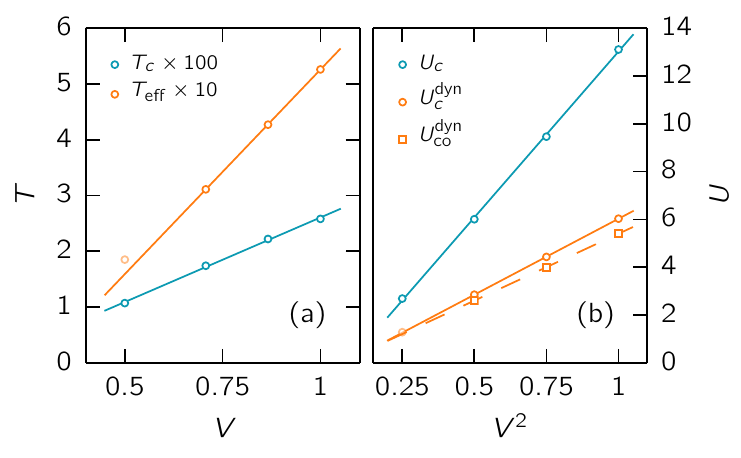}
  \caption{
    $V$-dependencies of (a) critical and effective temperatures (note the different rescaling) and (b)
    critical interactions for the periodic Anderson model with nearest-neighbor hybridization (circles). 
    Results for the equilibrium (blue) and the nonequilibrium case (orange). 
    Lines represent linear fits of the data.
    Right panel, squares: crossover interaction strength for the model with on-site hybridization.
    }
  \label{fig:UcTcV}
\end{figure}

The phase diagram is very much reminiscent of the phase diagram for the Mott transition in the Hubbard model. \citep{georges1993,rozenberg1994,potthoff2003b,pozgajcic2004}
In fact, based on DMFT calculations, \cite{held2000,held2000b} this similarity has been pointed out previously.  
In particular, \citet{held2000b} found an approximate $V^2$-scaling of the critical interaction for the PAM with nearest-neighbor hybridization at zero temperature.  
For the range of hybridization strengths considered here, this scaling is recovered within our calculations as is demonstrated with Fig.~\ref{fig:UcTcV}(b) (see the blue line).  
In addition, Fig.~\ref{fig:UcTcV}(a) (blue line) demonstrates that the critical temperature $T_{\rm c}$ scales approximately linearly with $V$.

\begin{figure}[b]
  \centering%
  \includegraphics[width=.66\columnwidth]{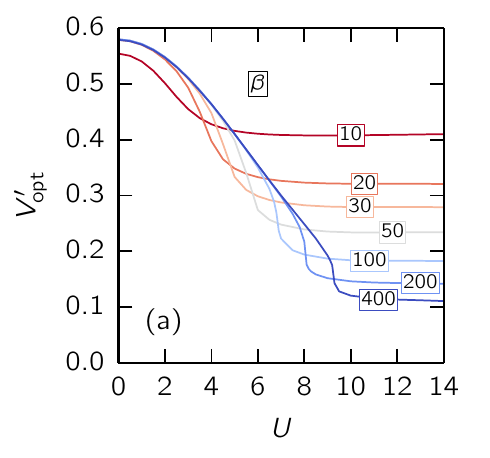}%
  \includegraphics[width=.33\columnwidth]{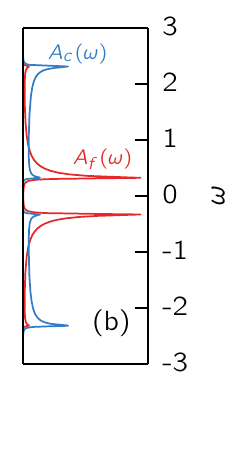}
  \caption{(a) $U$-dependence of the optimal variational parameter $V'_{\rm opt}$
    for the PAM with on-site hybridization $V^{2}=0.75$ and different
    inverse temperatures as indicated.  (b) Local spectral densities
    on the $f$ and $c$ orbitals for $U=0$ and zero temperature.}
  \label{fig:onsiteV}
\end{figure}

For the PAM with on-site hybridization [Fig.~\ref{fig:PAMs}(b)] we
have performed calculations for temperatures down to $T\geq 2.5 \times
10^{-3}$ (i.e., $\beta \leq 400$) but could not identify a coexistence
of different solutions.  This is demonstrated with
Fig.~\ref{fig:onsiteV}(a) which displays the optimal variational
parameter $V'_{\rm opt}$ as function of $U$ for fixed hybridization
$V^{2} = 0.75$ in the original system.  Starting from its
noninteracting value, $V'_{\rm opt}$ decreases monotonically with
increasing interaction and exhibits a steep slope close to an
inflection point.  It is tempting to anticipate this as a precursor
for some hysteresis behavior at even smaller temperatures $T$.
However, the position of the inflection point is proportional to $1/T$
which shifts the $T=0$ critical interaction to $U_{\rm c}\ra\infty$
as has been suggested by linearized DMFT. \citep{held2000b} This is
consistent with our expectation that the (paramagnetic) half-filled
PAM with an on-site hybridization has an insulating ground state and
crosses over from a hybridization band insulator at $U=0$ to a Kondo
insulator at strong $U$ without any quantum-critical point.  As an
illustration, Fig.~\ref{fig:onsiteV}(b) shows the $f$- and the
$c$-electron spectral densities at $U=0$.  For $V^{2} = 0.75$ the
hybridization band gap is clearly visible.

\section{Nonequilibrium}
\label{sec:nonequilibrium}
        
\subsection{Dynamical Mott transition}        
        
Let us now turn to the nonequilibrium Mott transition and start the discussion with the PAM with nearest-neighbor hybridization.
We first consider sudden quenches of the interaction strength at time $t=0$ from an essentially noninteracting initial state and ending at different final values $U_{\rm fin}$.
For convenience, we have chosen a small value $U_{\rm ini} = 0.1$ and an inverse temperature of $\beta=10$ for the initial state. 
As has been checked numerically, this is representative for a zero-temperature and noninteracting initial state, i.e., the chosen finite values for $U_{\rm ini}$ and $\beta$ do not have any significant impact on the subsequent nonequilibrium dynamics. 
Furthermore, we fix the nearest-neighbor hybridization at $V^{2}=0.75$ if not stated differently.

\begin{figure}[t]
\centering%
\includegraphics[width=\columnwidth]{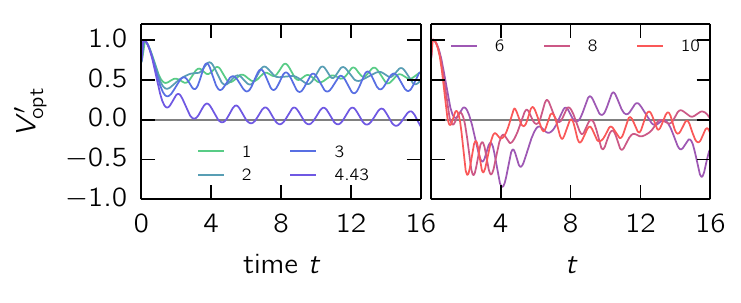}
\caption{
Time dependence of the optimal variational parameter $V'_{\rm opt}(t)$ for the periodic Anderson model with nearest-neighbor hybridization $V^2=0.75$ for quenches from $U_{\rm ini} = 0.1$ to different final interactions $U_{\rm fin}$ (see color labels). 
Left: $U_{\rm fin} \leq U_{\rm c}^{\rm dyn} \approx 4.43$. 
Right: $U_{\rm fin} > U_{\rm c}^{\rm dyn}$.
}
 \label{fig:V_offsite}
\end{figure}

Analyzing the results for different $U_{\rm fin}$, we find a dynamical Mott transition for quenches ending at an interaction $U_{\rm c}^{\rm dyn}\approx 4.43$.
This critical interaction sharply separates two distinct response regimes. 
Exemplary results for the time dependence of the optimal variational parameter $V'_{\rm opt}$ are shown in Fig.~\ref{fig:V_offsite}. 
For quenches to weak final interactions $U_{\rm fin} < U_{\rm c}^{\rm dyn}$, we observe a fast relaxation of $V'_{\rm opt}$ to a smaller but positive value within about two inverse hoppings. 
This is followed by moderate oscillations [see Fig.~\ref{fig:V_offsite} (left)]. 
Contrary, as shown in Fig.~\ref{fig:V_offsite} (right), for strong interactions $U_{\rm fin} > U_{\rm c}^{\rm dyn}$ the optimal variational parameter drops to
negative values. 
On top of the small and fast oscillations there are pronounced and slow beatings, the frequency of which increases with $U_{\rm fin}$. 
Right at the critical point $U_{\rm fin} = U_{\rm c}^{\rm dyn}$ the bath site dynamically decouples, i.e., for longer times the optimal variational parameter vanishes on average. 
As can be seen in Fig.~\ref{fig:V_offsite} (left), there are some residual oscillations around zero which are weak and regular but clearly present.
This is different from the behavior of $V'_{\rm opt}(t)$ in the Hubbard model where, for $U_{\rm fin} = U_{\rm c}^{\rm dyn}$, the bath site is found \citep{hofmann2016} to {\em exactly} decouple from the correlated site in the course of time.
The overall behavior in the different interaction regimes and at the critical interaction, however, is very similar to the Hubbard-model case.
We note that the critical interaction of the dynamical Mott transition, $U_{\rm c}^{\rm dyn} \approx 4.43$, is a bit more than a factor two smaller than the critical interaction, $U_{\rm c} \approx 9.46$ of the equilibrium Mott transition at $T=0$. 
The same ratio is found for the Hubbard model within the two-site DIA, \cite{hofmann2016} the Gutzwiller approach, \cite{schiro2010b} and within DMFT. \cite{eckstein2009b}

\subsection{Time-averaged quantities}

For the optimal variational parameter $V'_{\rm opt}$ and for derived physical observables, the double occupancy $\ev<n_\ua n_\da> \equiv \ev<
\nf{i\ua}\nf{i\da}>$, the total energy $E_{\rm tot}$ as well as the hybridization correlation $\ev<f^\dagger c> \equiv \ev<\cf{i\sigma}\ac{i\sigma}>$, we find a rather fast relaxation to some final values accompanied by regular oscillations.
Rather than on the precise dynamics, we therefore concentrate on the respective averages and define, for all time-dependent quantities $Q(t)$ of interest, the long-time average
\begin{equation}
  \overline Q = \lim_{t\ra\infty}\frac{1}{t}\int_0^t dt'\, Q(t')\,,
\end{equation}
and the variance
\begin{equation}
  {\Delta Q} = \left(\overline{(Q-\overline Q)^2}\right)^{\frac{1}{2}} \, . 
\label{eq:var}
\end{equation}

\begin{figure}[t]
\centering%
\includegraphics[width=\columnwidth]{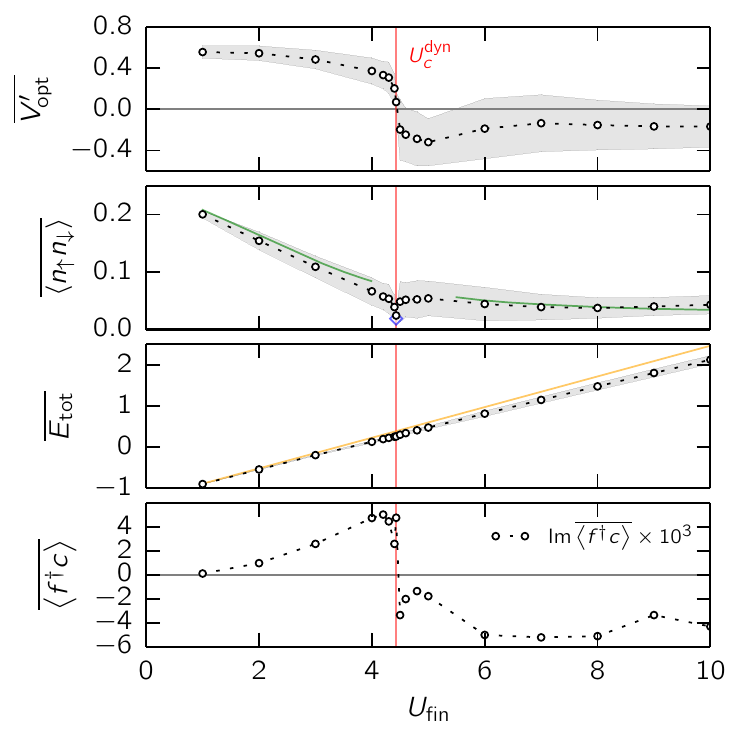}
\caption{
Long-time averages of the optimal variational parameter, of the double occupancy, of the total energy and of the imaginary part of the hybridization correlation as functions of $U_{\rm fin}$ for the PAM with nearest-neighbor hybridization $V^{2}=0.75$.
Black dashed lines serve as a guide to the eye.
The gray-shaded area indicates the variance of the oscillations around the average [see Eq.\ (\ref{eq:var})].
Red lines mark the critical interaction $U_{\rm c}^{\rm dyn}$. 
Green lines in the second panel: thermal values for the double occupancy as obtained from equilibrium two-site DIA calculations.
Blue diamond: the same but as obtained from a Hubbard-I calculation.
Yellow lines in the third panel: exact $U_{\rm fin}$-dependence of the total energy.
}
\label{fig:offsite_av}
\end{figure}

These are shown in Fig.~\ref{fig:offsite_av} as functions of $U_{\rm fin}$.
For weak $U_{\rm fin} < U_{\rm c}^{\rm dyn}$ the average of the optimal variational parameter $\overline{V'_{\rm opt}}$ (top panel) slowly decreases with increasing interaction, but then rapidly drops to zero.
This corresponds to the time-dependent decoupling of the bath site and defines the critical interaction. 
At and close to $U_{\rm c}^{\rm dyn}$, the variance ${\Delta V'_{\rm opt}}$ is at a minimum.
Beyond $U_{\rm c}^{\rm dyn}$, the sign of $\overline{V'_{\rm opt}}$ is negative.
One should note that the overall sign of $V'_{\rm opt}$ has no physical meaning. 
The stationary point of the self-energy functional and results for physical observables are actually invariant under a local U(1) gauge transformation $V'_{\rm opt} \to e^{i\varphi} V'_{\rm opt}$, and choosing a real parameter with $V'_{\rm opt} >0$ for weak $U$ only fixes the gauge.
Finally, for quenches ending at $U_{\rm fin} > U_{\rm c}^{\rm dyn}$ the absolute value of $\overline{V'_{\rm opt}}$ slowly decreases with $U_{\rm fin}$ but seems to saturate for $U_{\rm fin} \gtrsim 8$. 
For strong final interactions, this is accompanied by a considerably increased variance due to the slow beatings with large amplitude discussed above.

The fourth panel of Fig.~\ref{fig:offsite_av} shows the hybridization correlation. 
For the model with nearest-neighbor hybridization its real part is vanishing. 
The figure displays the imaginary part of the average only, i.e., $\mbox{Im} \overline{\ev<f^\dagger c>}$.
Note that this quantity is comparatively small and there are significant numerical errors, see the noise in the data for $U_{\rm fin}$ smaller but close to $U_{\rm c}^{\rm dyn}$, or the nonmonotonic trend for stronger $U_{\rm fin}$. 
Still, it is worth pointing out that the dynamical critical interaction is quite precisely characterized by a zero of the hybridization correlation while $\mbox{Im} \overline{\ev<f^\dagger c>} \ne 0$ for all $U_{\rm fin} < U_{\rm c}^{\rm dyn}$ and for all $U_{\rm fin} > U_{\rm c}^{\rm dyn}$.

In the initial noninteracting ground state, the double occupancy is given by $\ev<n_\ua n_\da>_0 = 0.25$. 
After the quench, it quickly relaxes within a few inverse hoppings and, due to the finite $U_{\rm fin}$, approaches a smaller value for longer times.
The time-averaged double occupancy (Fig.~\ref{fig:offsite_av}, second panel) almost linearly decreases with increasing $U_{\rm fin}$.
Interestingly, for strong interactions $5 \lesssim U_{\rm fin} \lesssim 10$ it remains almost constant at a small value of about $\overline{\ev<n_\ua n_\da>}=0.05$. 
Opposed to this, one would actually expect a larger double occupancy, at least in the infinite-$U_{\rm fin}$ limit.
The physical reason is that two electrons occupying the same $f$ orbital in the initial state at $t=0$ form a repulsively bound doublon which cannot decay on short time scales since the available phase space is strongly restricted by energy conservation.
This is well known for the Hubbard model in the strong-coupling limit \citep{eckstein2009b,strohmaier2010,hofmann2012,RP16} and also applies here. 
We must conclude that the final-state double occupancy is somewhat underestimated by the two-site DIA for strong $U_{\rm fin}$.
Right at the critical point $U_{\rm fin} = U_{\rm c}^{\rm dyn}$, the double occupancy rapidly drops to almost zero, and there are only small fluctuations around the time average. 
This result as well as the overall trend are again very much reminiscent of the findings for the dynamical Mott transition in the Hubbard model. \citep{schiro2010b,hofmann2016} 

The exact total energy at time $t=0^+$, immediately after the quench, is given by $E_{\rm tot} = E_{\rm kin}(0) + U_{\rm fin}/4$, i.e., by the expectation value of the final-state Hamiltonian in the noninteracting initial state.
Since the Hamiltonian is time independent after the quench, $E_{\rm tot}$ is constant.
Furthermore, it increases linearly with increasing $U_{\rm fin}$.
Within the two-site DIA, the total energy is computed from the self-energy and the approximate single-particle Green's function (see Ref.\ \onlinecite{hofmann2013}). 
The result is shown in the third panel of Fig.~\ref{fig:offsite_av} and can be seen to slightly deviate from the exact result (yellow line). 
It furthermore exhibits some moderate time-dependent oscillations (see gray-shaded area). 
In fact, while approximations constructed within the SFT strictly respect the conservation laws resulting from the invariance of the Hamiltonian under continuous transformation groups, i.e., conservation of particle number and spin, energy conservation can only be ensured within approximations resulting from a reference system with a continuum of bath degrees of freedom. 
Consequently, for the simple two-site reference system considered here, some violation of energy conservation must be tolerated. 

\subsection{Thermalization}        

One may compare the total energy in the long-time limit with its thermal expectation value in the grand-canonical ensemble at the same interaction strength $U_{\rm fin}$. 
Tentatively assuming that the system thermalizes, one must have $\overline{E_{\rm tot}} = E_{\rm tot}^{\rm eq}(T_{\rm eff})$ for the effective temperature $T_{\rm eff}$ of the thermal state.
This temperature can be computed within the two-site DIA by comparing the results shown in Fig.~\ref{fig:offsite_av} with those of corresponding equilibrium calculations for different temperatures.
$T_{\rm eff}$ almost linearly increases with increasing $U_{\rm fin}$ both, in the weak-coupling ($U_{\rm fin} < U_{\rm c}^{\rm dyn}$) and in the strong-coupling regime ($U_{\rm fin} > U_{\rm c}^{\rm dyn}$), but in the latter the slope $\partial T_{\rm eff}/\partial U_{\rm fin}$ turns out to be about an order of magnitude larger.

To check whether or not the system indeed thermalizes, we compare the long-time average $\overline{\ev<n_\ua n_\da>}$ of the double occupancy after the quench with the value of the double occupancy obtained from an equilibrium two-site DIA calculation at interaction $U_{\rm fin}$ and temperature $T_{\rm eff}$, displayed as green lines in the second panel of Fig.\ \ref{fig:offsite_av}.
There is reasonable agreement for interaction strengths $U_{\rm fin}$ which are not too close to $U_{\rm c}^{\rm dyn}$.

Right at the critical point, however, agreement with a thermal double occupancy is only found when the comparison is done with the equilibrium state that is obtained within the Hubbard-I approximation, i.e., if the variational parameter $V'_{\rm opt}$ is {\em ad hoc} set to zero or for a reference system without any bath site at all.
The effective temperature estimated in this way is $T_{\rm eff} \approx 0.43$. 
The choice $V'_{\rm opt} = 0$ in fact always gives a stationary point of the self-energy functional but here corresponds to a thermal state which is metastable only since $V'_{\rm opt} > 0$ in the stable thermal state at any $U_{\rm fin}$ and $T>0$.
On the other hand, $V'_{\rm opt} = 0$ is consistent with the fact that the bath site dynamically decouples in the long-time limit.

We conclude that the system seems to approach a thermal (or metastable thermal) state in the long-time limit for the different interaction regimes discussed above but actually one would like to rely such a characterization on further observables. 
Here, the momentum-distribution function suggests itself since this can easily be derived from the Green's function and, as a quantity defined in reciprocal space, is complementary to the double occupancy.
Unfortunately, the distribution function exhibits a strongly oscillatory behavior which we ascribe to the small reference system and which does not permit meaningful quantitative analysis.

Opposed to the Hubbard model, the equilibrium Mott transition in the periodic Anderson model with nearest-neighbor hybridization is orbital selective since the one-particle gap opens in the $f$ spectral function only (see Fig.\ \ref{fig:offsiteA}). 
For essentially the same reason, also the time-dependent Mott transition is orbital selective: 
Consider the quench to $U_{\rm fin} = U_{\rm c}^{\rm dyn}$.
After the bath site has decoupled from the correlated site in the reference system, i.e., in the long-time limit, the spectral function in the metastable thermal state is obtained from the Green's function in Eq.\ (\ref{eq:Gsftk}) by summing over $k$ and taking the imaginary part, where $\Sigma'_{\rm opt}(\omega)$ is the one-pole Hubbard-I self-energy. 
The spectrum is the same as the one shown in Fig.\ \ref{fig:offsiteA} at $U=U_{\rm c}$, i.e., for the interaction strength where the bath site has just decoupled. 
Here the $f$ spectral function exhibits the Mott-Hubbard gap while the $c$ spectral function is gapless.
At finite temperatures there is of course no clear-cut distinction between metallic and insulating behavior but the above-estimated effective temperature $T_{\rm eff} \approx 0.43$ is still much smaller than the Mott-Hubbard gap.

\subsection{$V$ dependence}        

So far we have discussed results for a hybridization fixed at $V^{2}=0.75$.
Repeating the calculations for other values of $V$, we find the essentially same physics, i.e., a dynamical Mott transition. 
The critical interaction becomes $V$-dependent. 
As is demonstrated with Fig.~\ref{fig:UcTcV}(b) (see orange symbols and lines), we find a $V^2$-scaling of $U_{\rm c}^{\rm dyn}$.
This is very much reminiscent of the approximate $V^{2}$-dependence of the critical interaction for the zero-temperature Mott transition which has been found numerically and which could also be derived by exploiting the system's properties right at the critical point. \citep{held2000b} 
The scaling thus indicates that both effects are related and that both, the equilibrium and the nonequilibrium Mott transition, share the same critical behavior to some extent. 
In fact, within the two-site DIA and as discussed above, both transitions are basically characterized by the decoupling of the bath site in the reference system.

Another relation between the equilibrium and the nonequilibrium transition is the fact that the ratio between the respective critical interactions is, independent of $V$, always roughly given by $U_{\rm c}(T=0) / U_{\rm c}^{\rm dyn} \approx 2$, see Fig.~\ref{fig:UcTcV}(b).

Finally, also the $V$ dependencies of the critical temperature $T_{c}$ for the equilibrium transition and of the effective temperature $T_{\rm eff}$ for the nonequilibrium transition are the same. 
Both temperatures scale approximately linearly with $V$ as can be seen in Fig.~\ref{fig:UcTcV}(a).

\subsection{Ramping the interaction}

To further investigate the relation between the Mott transitions in and out of equilibrium, let us discuss the real-time dynamics after an interaction ramp with different ramp durations $\Delta t_{\rm ramp}$.
The considered ramp profile is 
\begin{equation}
U(t) = U_{\rm ini} + (U_{\rm fin} - U_{\rm ini}) \frac{1-\cos(\pi t/\Delta t_{\rm ramp})}{2} \; . 
\label{eq:ramp}
\end{equation}  
As above we choose $U_{\rm ini}=0.1$ and consider ramps to different $U_{\rm fin}$.
The hybridization is again fixed at $V^{2} = 0.75$.
For $\Delta t_{\rm ramp} = 0$ we trivially recover the above-discussed quench dynamics where the dynamical Mott transition is obtained for $U_{\rm fin} = U_{\rm c}^{\rm dyn}$.
In the limit $\Delta t_{\rm ramp}\to \infty$, on the other hand, the system is forced to evolve in an adiabatic process in the equilibrium phase diagram and, for $U_{\rm fin} > U_{\rm c}(T=0)$, to cross the equilibrium phase boundary (cf. Fig.~\ref{fig:pds}). 
Note that, as we start from an initial state which essentially can be considered as a zero-temperature state, an adiabatic process will also result in a zero-temperature final state.

\begin{figure}[t]
  \centering%
  \includegraphics[width=\columnwidth]{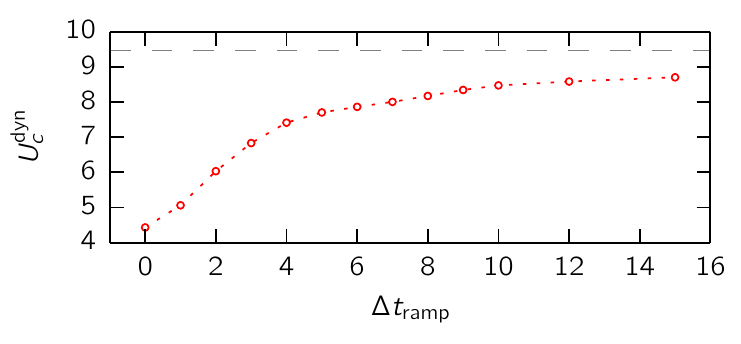}
  \caption{Critical interaction $U_{\rm c}^{\rm dyn}$ of the time-dependent Mott transition as a
    function of the ramp time $\Delta t_{\rm ramp}$ for the periodic Anderson model with nearest-neighbor hybridization
    $V^{2}=0.75$. Gray dashed line: equilibrium critical interaction
    $U_{\rm c}^{\rm dyn} \approx 9.46$ at zero temperature.}
  \label{fig:Uc_dtramp}
\end{figure}

Due to numerical limitations, we study ramps with durations $\Delta t_{\rm ramp} \leq15$. 
Varying the final interaction, we qualitatively observe the same behavior as in the quench case: 
There are two distinct response regimes which are sharply separated by a critical interaction $U_{\rm c}^{\rm dyn}$ which depends on $\Delta t_{\rm ramp}$.
For ramps with $\Delta t_{\rm ramp}\gtrsim 3$, the small regular oscillations observed for all quantities in the post-quench dynamics almost completely fade out. 
For weak interactions, the optimal variational parameter, for example, relaxes to a constant and positive value on a time scale of the order of $\Delta t_{\rm ramp}$.
Contrary, for strong interactions it performs slow collapse-and-revival oscillations around some negative value. 
At the critical interaction, the bath site exactly decouples in the long-time limit, $V'_{\rm opt}(t) \to 0$.

The $\Delta t_{\rm ramp}$ dependence of the critical interaction is shown in Fig.~\ref{fig:Uc_dtramp}.
It smoothly bridges the two limits $\Delta t_{\rm ramp} = 0$ and $\Delta t_{\rm ramp} = \infty$. 
In particular, for slower and slower ramps $U_{\rm c}^{\rm dyn}$ monotonically increases and appears to converge toward the equilibrium critical interaction $U_{\rm c}$ at zero temperature. 
This strongly suggests that both the equilibrium Mott transition and the nonequilibrium critical behavior are smoothly connected. 
A similar dependence of the critical interaction on the ramp time has been confirmed also for the Hubbard model within
two-site DIA \citep{hofmann2016} and the Gutzwiller method. \citep{sandri2012}

\subsection{On-site hybridization}

\begin{figure}[t]
\centering%
\includegraphics[width=\columnwidth]{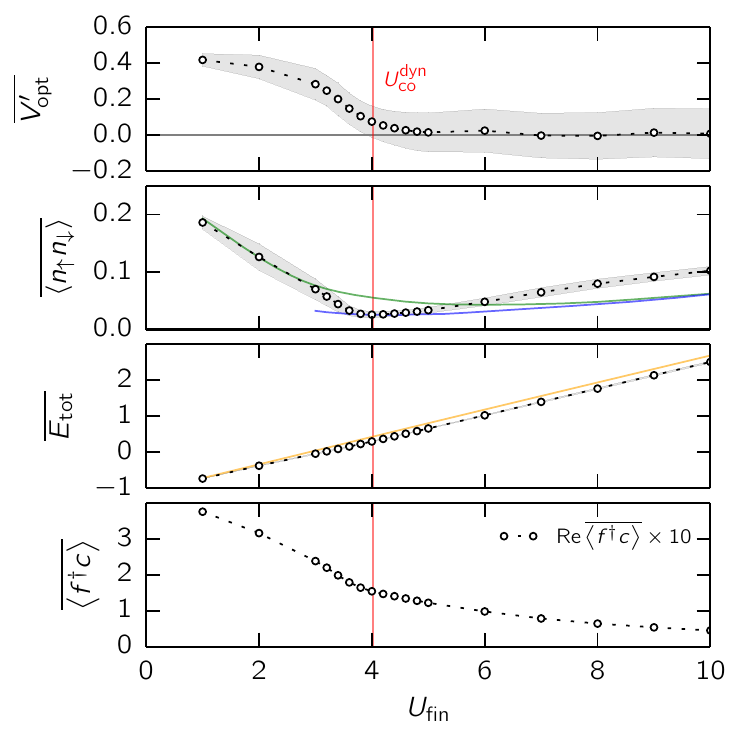}
\caption{
The same as Fig.\ \ref{fig:offsite_av} but for on-site hybridization ($V^{2}=0.75$).  
Red lines: ``crossover'' interaction $U_{\rm co}^{\rm dyn}$. 
}
\label{fig:onsite_av}
\end{figure}

There is no Mott transition in the equilibrium phase diagram of the periodic Anderson model with on-site hybridization [cf.\ Fig.~\subref*{fig:PAMonsite}].
We therefore expect that also the time-dependent Mott transition is absent in the same model.
The results of corresponding calculations are shown Fig.~\ref{fig:onsite_av}. 
For a direct comparison with the off-site-hybridization case, Fig.~\ref{fig:offsite_av}, we again consider the real-time dynamics following an interaction quench for exactly the same parameters.

At weak interaction $U_{\rm fin}$, we essentially find the same features as described for the case of nearest-neighbor hybridization. 
In particular, the trend of all time-averaged parameters and observables is almost the same. 
An expected exception is the hybridization correlation.
In the on-site case $\Im\overline{\ev<f^\dagger c>}\equiv 0$, and only the real part is plotted. 
It is positive, decreases monotonically with $U_{\rm fin}$, and shows a completely regular behavior.
Opposed to the off-site case, there is no sharp transition in the $U_{\rm fin}$-dependence of all quantities.
This not only applies to the time-averaged values but also to the $U_{\rm fin}$-dependence of the respective real-time dynamics as has been checked carefully.

The time-averaged optimal hybridization $\overline{V'_{\rm opt}}$ monotonically decreases and quickly approaches very small values for $U_{\rm fin} \gtrsim 5$.
This is accompanied by pronounced regular oscillations around the average value, see the gray-shaded area in the top panel of Fig.~\ref{fig:onsite_av}. 
Opposed to the off-site case, however, there is no sign change $\overline{V'_{\rm opt}}$ and no indication for a dynamical Mott transition at a finite $U_{\rm fin}$.

For the double occupancy, after a quick initial drop, we find almost time-independent values for all interaction strengths $U_{\rm fin}$, as can be seen from the small fluctuations displayed in Fig.~\ref{fig:onsite_av}.  
As a function of $U_{\rm fin}$, the time-averaged double occupancy decreases.
It is at a minimum for $U_{\rm fin} \approx 4$, and then increases again due to the stability of repulsively bound doublons in the limit of strong interaction, as already discussed for the case of nearest-neighbor hybridization.

We refer to $U_{\rm fin} = U_{\rm co}^{\rm dyn} \approx 4$ as the ``crossover interaction''. 
This is marked by the red lines in Fig.\ \ref{fig:onsite_av}.
Although there is no dynamical Mott transition in the Anderson model with on-site hybridization, $U_{\rm co}^{\rm dyn}$ marks the minimum of the time-averaged double occupancy, and the inflection point of the $\overline{V'_{\rm opt}}$ as well as of 
$\overline{\ev<\cf{}\ac{}>}$ as function of $U_{\rm fin}$. 

Conservation of the total energy is respected to a better degree for the case of on-site hybridization (see third panel of Fig.~\ref{fig:onsite_av}).
Assuming that the system thermalizes, we compute the effective temperature via $\overline{E_{\rm tot}} = E_{\rm tot}^{\rm eq}(T_{\rm eff})$. 
It turns out that $T_{\rm eff}$ roughly scales as $U_{\rm fin}^{2}$ over the entire range of interaction strengths considered (not shown).
This differs from the on-site case, where two distinct interaction regimes with linear dependencies on $U_{\rm fin}$ but largely different slopes could be identified.

The time-averaged double occupancy can be compared with the respective equilibrium value using the effective temperature for the thermal average (green lines in Fig.\ \ref{fig:onsite_av}, second panel). 
For weak interactions, up to $U_{\rm fin} \lesssim 3$, the agreement is almost perfect. 
For strong $U_{\rm fin}$, the time average exceeds the thermal value.
In the vicinity of the crossover interaction $U_{\rm co}^{\rm dyn}$, the time-averaged double occupancy agrees well with the thermal double occupancy in the metastable state obtained by {\em ad hoc} setting the variational parameter to zero (blue line). 
This is remarkable since, opposed to the case of the model with nearest-neighbor hybridization, the bath site does {\em not} decouple dynamically. 

We have also performed calculations for different values of the hybridization $V$. 
The overall behavior of the optimal variational parameter, of the double occupancy, total energy and hybridization correlation does not change. 
As can be seen in Fig.~\ref{fig:UcTcV}(b) (squares), the crossover interaction follows a $V^{2}$-trend, very similar to the critical interaction for the dynamical Mott transition in the model with nearest-neighbor hybridization (circles).

\section{Conclusions}
\label{sec:conclusions}

The recently developed generalization \cite{hofmann2013} of the self-energy functional theory to systems far from equilibrium has been applied to study the time-dependent Mott transition in a lattice model with two-orbitals per unit cell. 
Here, the critical interaction for the dynamical but also for the equilibrium Mott transition are expected to depend on further parameters such that their mutual relation can be studied by scanning the parameter space. 
The simplest realization of this idea consists in an application of the nonequilibrium two-site dynamical impurity approximation (two-site DIA) to the periodic Anderson model. 
Within the dynamical mean-field theory or in the limit of infinite spatial dimensions, a variant of this model, with a nearest-neighbor hybridization between $f$ and $c$ orbitals, exhibits an orbital-selective Mott transition at half filling as has been demonstrated in several earlier studies. \cite{huscroft1999,held2000,held2000b,vandongen2001}

Let us summarize the main findings of the present work, starting with the equilibrium Mott transition:
With the two-site DIA one can successfully reproduce the previous results for the zero-temperature Mott transition and furthermore compute the whole phase diagram in the $T$-$U$ plane at half filling. 
The phenomenology turns out as very similar to the Mott transition in the single-orbital Hubbard model \cite{georges1996} with a first-order transition at finite temperature at a critical interaction $U_{\rm c}(T)$ for temperatures $T<T_{\rm c}$. 
In the Anderson model with nearest-neighbor hybridization, we find $U_{\rm c} \equiv U_{\rm c}(T=0)$ to approximately scale as $V^{2}$ and $T_{\rm c}$ to scale linear with the hybridization strength $V$. 
Using the linearized DMFT \cite{held2000b,bulla2000} and right at $U_{\rm c}$, there is in fact an approximate mapping of the model onto the Hubbard model which explains the approximate $V^2$-scaling of $U_{\rm c}$.
Contrary, the application of the two-site DIA to the model with on-site hybridization does not yield a Mott transition which is the expected result as this model variant should be a band or Kondo insulator at any interaction strength.
The absence of the Kondo physics in the Anderson model with nearest-neighbor hybridization can be explained \cite{vandongen2001} by the fact that the $k$-dependent hybridization strength just vanishes at the Fermi surface of the noninteracting system. 

The application of the nonequilibrium two-site DIA to study the real-time dynamics initiated by an interaction quench or ramp reveals that there is a time-dependent Mott transition in the Anderson model with nearest-neighbor hybridization in fact. 
There are many characteristics which are reminiscent of the dynamical Mott transition in the Hubbard model, \cite{eckstein2009b,eckstein2010b,schiro2010b,hofmann2016} such as 
two distinct response regimes, characterized by either a quick relaxation of the optimal variational parameter toward an almost constant value for weak $U$ or slow, pronounced collapse-and-revival oscillations for strong interactions. 
Both regimes are sharply separated by a critical interaction $U_{\rm c}^{\rm dyn}$ where the bath site in the DIA reference system essentially decouples from the correlated site in the course of time. 
Within the two-site DIA, this time-dependent transition is very similar to the zero-temperature Mott transition which is also characterized by a (quasistatic) decoupling of the bath site.
The final state after the quench to $U_{\rm c}^{\rm dyn}$ can be described as a thermal state with effective temperature $T_{\rm eff}$ but is metastable as there is another equilibrium stationary point of the self-energy functional with lower grand potential.
Opposed to the Hubbard model, the time-dependent Mott transition is orbital selective, i.e., only the $f$ spectral function of the metastable thermal state after the quench develops the Mott gap.

We could further demonstrate that the equilibrium and the nonequilibrium Mott transition are closely related.
First of all, for the model variant with on-site hybridization where there is no transition in the thermodynamical state space, we also find the absence of a time-dependent Mott transition.
It has been argued, \cite{vandongen2001} that the equilibrium Mott transition in the Anderson model with nearest-neighbor hybridization is fragile as a small ``perturbation'' by a finite on-site hybridization will turn the transition into a crossover. 
This fragile character seems to be proliferated to the nonequilibrium transition: 
For the on-site case, we in fact find a smooth dynamical crossover rather than sharp transition in the post-quench dynamics.
At this point, one may also mention a related study of the Hubbard model \cite{schiro2010b,schiro2011} where the disappearance of time-dependent critical behavior (accompanying the disappearance of criticality in the thermodynamical sense) upon doping the system away from half filling has been reported.

There is more evidence for a close relation between the equilibrium and the nonequilibrium Mott transition: 
The critical interaction for the Mott transition depends on the hybridization strength. 
Our calculations show that for both, the equilibrium and the nonequilibrium case, the critical interaction scales approximately linearly with $V^{2}$ for the $V$-range studied here. 
Interestingly, in all cases known so far, \cite{eckstein2009b,eckstein2010b,schiro2010b,hofmann2016} including the results for different $V$ of the present work, the ratio between $U_{\rm c}$ and $U_{\rm c}^{\rm dyn}$ is roughly given by two.
Furthermore, the characteristic temperature $T_{\rm eff}$ for the nonequilibrium transition has a different (namely linear) $V$-dependence which, however, is again the same as the $V$-dependence of the critical temperature $T_{\rm c}$ in the equilibrium case.
Finally, a straightforward link between the equilibrium and the nonequilibrium transition emerges in studies where the interaction is ramped to $U_{\rm fin}$ within a finite time $\Delta t_{\rm ramp}$. 
In agreement with previous studies of the Hubbard model, \cite{schiro2010b,hofmann2016} we find that the critical interaction smoothly interpolates between the result for the sudden quench $U_{\rm c}^{\rm dyn} = U_{\rm c}^{\rm dyn}(\Delta t_{\rm ramp}=0)$ and an quasistatic, adiabatic thermodynamical process $U_{\rm c}(T=0) = U_{\rm c}^{\rm dyn}(\Delta t_{\rm ramp} \to \infty)$.
Further and more systematic studies along these lines but with longer propagation times would be necessary to extract critical exponents of the (quantum) Kibble-Zurek mechanism. \cite{dziarmaga2010,dziarmaga2016}

Concluding, there are apparent similarities and close links between the two types of Mott transitions. 
One may speculate that the intimate relation between the equilibrium and the nonequilibrium Mott transition is due to the same ``effective theory'' which is able to describe the critical behavior in both cases, such as the effective, simplified physical pictures that are provided by the linearized DMFT \cite{bulla2000} or the projective self-consistent method \cite{Moll95} for the zero-temperature Mott transition only.
Work along these lines appears as promising.
The two-site DIA itself can already be regarded as an effective low-energy theory in this respect since it focusses on the time-dependent hybridization of the bath site at $\omega=0$. 
An important open question is to what extent the physics found within this approach is representative for the physics of the full nonequilibrium DMFT. 
Future studies using the DIA but involving more bath sites or more extensive nonequilibrium DMFT studies based on a nonperturbative solver are required, for the single- and for multi-orbital Hubbard-type models.

\acknowledgments

We would like to thank K.\ Balzer, M.\ Eckstein and M.\ Sayad for numerous instructive discussions and computer codes.  
Financial support of this work by the Deutsche Forschungsgemeinschaft within the Sonderforschungsbereich 925 (project B5) is gratefully acknowledged.

\bibliographystyle{apsrev4-1}

\begin{thebibliography}{61}%
\makeatletter
\providecommand \@ifxundefined [1]{%
 \@ifx{#1\undefined}
}%
\providecommand \@ifnum [1]{%
 \ifnum #1\expandafter \@firstoftwo
 \else \expandafter \@secondoftwo
 \fi
}%
\providecommand \@ifx [1]{%
 \ifx #1\expandafter \@firstoftwo
 \else \expandafter \@secondoftwo
 \fi
}%
\providecommand \natexlab [1]{#1}%
\providecommand \enquote  [1]{``#1''}%
\providecommand \bibnamefont  [1]{#1}%
\providecommand \bibfnamefont [1]{#1}%
\providecommand \citenamefont [1]{#1}%
\providecommand \href@noop [0]{\@secondoftwo}%
\providecommand \href [0]{\begingroup \@sanitize@url \@href}%
\providecommand \@href[1]{\@@startlink{#1}\@@href}%
\providecommand \@@href[1]{\endgroup#1\@@endlink}%
\providecommand \@sanitize@url [0]{\catcode `\\12\catcode `\$12\catcode
  `\&12\catcode `\#12\catcode `\^12\catcode `\_12\catcode `\%12\relax}%
\providecommand \@@startlink[1]{}%
\providecommand \@@endlink[0]{}%
\providecommand \url  [0]{\begingroup\@sanitize@url \@url }%
\providecommand \@url [1]{\endgroup\@href {#1}{\urlprefix }}%
\providecommand \urlprefix  [0]{URL }%
\providecommand \Eprint [0]{\href }%
\providecommand \doibase [0]{http://dx.doi.org/}%
\providecommand \selectlanguage [0]{\@gobble}%
\providecommand \bibinfo  [0]{\@secondoftwo}%
\providecommand \bibfield  [0]{\@secondoftwo}%
\providecommand \translation [1]{[#1]}%
\providecommand \BibitemOpen [0]{}%
\providecommand \bibitemStop [0]{}%
\providecommand \bibitemNoStop [0]{.\EOS\space}%
\providecommand \EOS [0]{\spacefactor3000\relax}%
\providecommand \BibitemShut  [1]{\csname bibitem#1\endcsname}%
\let\auto@bib@innerbib\@empty
\bibitem [{\citenamefont {Mott}(1949)}]{mott1949}%
  \BibitemOpen
  \bibfield  {author} {\bibinfo {author} {\bibfnamefont {N.~F.}\ \bibnamefont
  {Mott}},\ }\href@noop {} {\bibfield  {journal} {\bibinfo  {journal} {Proc.
  Phys. Soc. (London)}\ }\textbf {\bibinfo {volume} {A62}},\ \bibinfo {pages}
  {416} (\bibinfo {year} {1949})}\BibitemShut {NoStop}%
\bibitem [{\citenamefont {Mott}(1990)}]{mott1990}%
  \BibitemOpen
  \bibfield  {author} {\bibinfo {author} {\bibfnamefont {N.}~\bibnamefont
  {Mott}},\ }\href@noop {} {\emph {\bibinfo {title} {Metal-Insulator
  Transitions}}},\ \bibinfo {edition} {2nd}\ ed.\ (\bibinfo  {publisher}
  {Taylor \& Francis},\ \bibinfo {address} {London},\ \bibinfo {year}
  {1990})\BibitemShut {NoStop}%
\bibitem [{\citenamefont {Gebhard}(1997)}]{gebhard2010}%
  \BibitemOpen
  \bibfield  {author} {\bibinfo {author} {\bibfnamefont {F.}~\bibnamefont
  {Gebhard}},\ }\href@noop {} {\emph {\bibinfo {title} {The Mott
  Metal-Insulator Transition}}}\ (\bibinfo  {publisher} {Springer},\ \bibinfo
  {address} {Berlin},\ \bibinfo {year} {1997})\BibitemShut {NoStop}%
\bibitem [{\citenamefont {Georges}\ \emph {et~al.}(1996)\citenamefont
  {Georges}, \citenamefont {Kotliar}, \citenamefont {Krauth},\ and\
  \citenamefont {Rozenberg}}]{georges1996}%
  \BibitemOpen
  \bibfield  {author} {\bibinfo {author} {\bibfnamefont {A.}~\bibnamefont
  {Georges}}, \bibinfo {author} {\bibfnamefont {G.}~\bibnamefont {Kotliar}},
  \bibinfo {author} {\bibfnamefont {W.}~\bibnamefont {Krauth}}, \ and\ \bibinfo
  {author} {\bibfnamefont {M.~J.}\ \bibnamefont {Rozenberg}},\ }\href@noop {}
  {\bibfield  {journal} {\bibinfo  {journal} {Rev. Mod. Phys.}\ }\textbf
  {\bibinfo {volume} {68}},\ \bibinfo {pages} {13} (\bibinfo {year}
  {1996})}\BibitemShut {NoStop}%
\bibitem [{\citenamefont {Kotliar}\ and\ \citenamefont
  {Vollhardt}(2004)}]{kotliar2004}%
  \BibitemOpen
  \bibfield  {author} {\bibinfo {author} {\bibfnamefont {G.}~\bibnamefont
  {Kotliar}}\ and\ \bibinfo {author} {\bibfnamefont {D.}~\bibnamefont
  {Vollhardt}},\ }\href@noop {} {\bibfield  {journal} {\bibinfo  {journal}
  {Physics Today}\ }\textbf {\bibinfo {volume} {57}},\ \bibinfo {pages} {53}
  (\bibinfo {year} {2004})}\BibitemShut {NoStop}%
\bibitem [{\citenamefont {Eckstein}\ \emph {et~al.}(2009)\citenamefont
  {Eckstein}, \citenamefont {Kollar},\ and\ \citenamefont
  {Werner}}]{eckstein2009b}%
  \BibitemOpen
  \bibfield  {author} {\bibinfo {author} {\bibfnamefont {M.}~\bibnamefont
  {Eckstein}}, \bibinfo {author} {\bibfnamefont {M.}~\bibnamefont {Kollar}}, \
  and\ \bibinfo {author} {\bibfnamefont {P.}~\bibnamefont {Werner}},\
  }\href@noop {} {\bibfield  {journal} {\bibinfo  {journal} {Phys. Rev. Lett.}\
  }\textbf {\bibinfo {volume} {103}},\ \bibinfo {pages} {056403} (\bibinfo
  {year} {2009})}\BibitemShut {NoStop}%
\bibitem [{\citenamefont {Schir\'o}\ and\ \citenamefont
  {Fabrizio}(2010)}]{schiro2010b}%
  \BibitemOpen
  \bibfield  {author} {\bibinfo {author} {\bibfnamefont {M.}~\bibnamefont
  {Schir\'o}}\ and\ \bibinfo {author} {\bibfnamefont {M.}~\bibnamefont
  {Fabrizio}},\ }\href@noop {} {\bibfield  {journal} {\bibinfo  {journal}
  {Phys. Rev. Lett.}\ }\textbf {\bibinfo {volume} {105}},\ \bibinfo {pages}
  {076401} (\bibinfo {year} {2010})}\BibitemShut {NoStop}%
\bibitem [{\citenamefont {Hamerla}\ and\ \citenamefont
  {Uhrig}(2013)}]{hamerla2013}%
  \BibitemOpen
  \bibfield  {author} {\bibinfo {author} {\bibfnamefont {S.~A.}\ \bibnamefont
  {Hamerla}}\ and\ \bibinfo {author} {\bibfnamefont {G.~S.}\ \bibnamefont
  {Uhrig}},\ }\href@noop {} {\bibfield  {journal} {\bibinfo  {journal} {Phys.
  Rev. B}\ }\textbf {\bibinfo {volume} {87}},\ \bibinfo {pages} {064304}
  (\bibinfo {year} {2013})}\BibitemShut {NoStop}%
\bibitem [{\citenamefont {Hamerla}\ and\ \citenamefont
  {Uhrig}(2014)}]{hamerla2014}%
  \BibitemOpen
  \bibfield  {author} {\bibinfo {author} {\bibfnamefont {S.~A.}\ \bibnamefont
  {Hamerla}}\ and\ \bibinfo {author} {\bibfnamefont {G.~S.}\ \bibnamefont
  {Uhrig}},\ }\href@noop {} {\bibfield  {journal} {\bibinfo  {journal} {Phys.
  Rev. B}\ }\textbf {\bibinfo {volume} {89}},\ \bibinfo {pages} {104301}
  (\bibinfo {year} {2014})}\BibitemShut {NoStop}%
\bibitem [{\citenamefont {Esslinger}(2010)}]{esslinger2010}%
  \BibitemOpen
  \bibfield  {author} {\bibinfo {author} {\bibfnamefont {T.}~\bibnamefont
  {Esslinger}},\ }\href@noop {} {\bibfield  {journal} {\bibinfo  {journal}
  {Annu. Rev. Condens. Matter Phys.}\ }\textbf {\bibinfo {volume} {1}},\
  \bibinfo {pages} {129} (\bibinfo {year} {2010})}\BibitemShut {NoStop}%
\bibitem [{\citenamefont {Bloch}\ \emph {et~al.}(2012)\citenamefont {Bloch},
  \citenamefont {Dalibard},\ and\ \citenamefont {Nascimb\`{e}ne}}]{bloch2012}%
  \BibitemOpen
  \bibfield  {author} {\bibinfo {author} {\bibfnamefont {I.}~\bibnamefont
  {Bloch}}, \bibinfo {author} {\bibfnamefont {J.}~\bibnamefont {Dalibard}}, \
  and\ \bibinfo {author} {\bibfnamefont {S.}~\bibnamefont {Nascimb\`{e}ne}},\
  }\href@noop {} {\bibfield  {journal} {\bibinfo  {journal} {Nat. Phys.}\
  }\textbf {\bibinfo {volume} {8}},\ \bibinfo {pages} {267} (\bibinfo {year}
  {2012})}\BibitemShut {NoStop}%
\bibitem [{\citenamefont {Lewenstein}\ \emph {et~al.}(2012)\citenamefont
  {Lewenstein}, \citenamefont {Sanpera},\ and\ \citenamefont
  {Ahufinger}}]{lewenstein2012}%
  \BibitemOpen
  \bibfield  {author} {\bibinfo {author} {\bibfnamefont {M.}~\bibnamefont
  {Lewenstein}}, \bibinfo {author} {\bibfnamefont {A.}~\bibnamefont {Sanpera}},
  \ and\ \bibinfo {author} {\bibfnamefont {V.}~\bibnamefont {Ahufinger}},\
  }\href@noop {} {\emph {\bibinfo {title} {Ultracold Atoms in Optical Lattices:
  Simulating quantum many-body systems}}},\ \bibinfo {edition} {1st}\ ed.\
  (\bibinfo  {publisher} {Oxford University Press},\ \bibinfo {address}
  {Oxford},\ \bibinfo {year} {2012})\BibitemShut {NoStop}%
\bibitem [{\citenamefont {Srednicki}(1994)}]{srednicki1994}%
  \BibitemOpen
  \bibfield  {author} {\bibinfo {author} {\bibfnamefont {M.}~\bibnamefont
  {Srednicki}},\ }\href@noop {} {\bibfield  {journal} {\bibinfo  {journal}
  {Phys. Rev. E}\ }\textbf {\bibinfo {volume} {50}},\ \bibinfo {pages} {888}
  (\bibinfo {year} {1994})}\BibitemShut {NoStop}%
\bibitem [{\citenamefont {Deutsch}(1991)}]{deutsch1991}%
  \BibitemOpen
  \bibfield  {author} {\bibinfo {author} {\bibfnamefont {J.~M.}\ \bibnamefont
  {Deutsch}},\ }\href@noop {} {\bibfield  {journal} {\bibinfo  {journal} {Phys.
  Rev. A}\ }\textbf {\bibinfo {volume} {43}},\ \bibinfo {pages} {2046}
  (\bibinfo {year} {1991})}\BibitemShut {NoStop}%
\bibitem [{\citenamefont {Rigol}\ \emph {et~al.}(2008)\citenamefont {Rigol},
  \citenamefont {Dunjko},\ and\ \citenamefont {Olshanii}}]{rigol2008}%
  \BibitemOpen
  \bibfield  {author} {\bibinfo {author} {\bibfnamefont {M.}~\bibnamefont
  {Rigol}}, \bibinfo {author} {\bibfnamefont {V.}~\bibnamefont {Dunjko}}, \
  and\ \bibinfo {author} {\bibfnamefont {M.}~\bibnamefont {Olshanii}},\
  }\href@noop {} {\bibfield  {journal} {\bibinfo  {journal} {Nature}\ }\textbf
  {\bibinfo {volume} {452}},\ \bibinfo {pages} {854} (\bibinfo {year}
  {2008})}\BibitemShut {NoStop}%
\bibitem [{\citenamefont {Berges}\ \emph {et~al.}(2004)\citenamefont {Berges},
  \citenamefont {Borsányi},\ and\ \citenamefont {Wetterich}}]{berges2004}%
  \BibitemOpen
  \bibfield  {author} {\bibinfo {author} {\bibfnamefont {J.}~\bibnamefont
  {Berges}}, \bibinfo {author} {\bibfnamefont {S.}~\bibnamefont {Borsányi}}, \
  and\ \bibinfo {author} {\bibfnamefont {C.}~\bibnamefont {Wetterich}},\
  }\href@noop {} {\bibfield  {journal} {\bibinfo  {journal} {Phys. Rev. Lett.}\
  }\textbf {\bibinfo {volume} {93}},\ \bibinfo {pages} {142002} (\bibinfo
  {year} {2004})}\BibitemShut {NoStop}%
\bibitem [{\citenamefont {Moeckel}\ and\ \citenamefont
  {Kehrein}(2008)}]{moeckel2008}%
  \BibitemOpen
  \bibfield  {author} {\bibinfo {author} {\bibfnamefont {M.}~\bibnamefont
  {Moeckel}}\ and\ \bibinfo {author} {\bibfnamefont {S.}~\bibnamefont
  {Kehrein}},\ }\href@noop {} {\bibfield  {journal} {\bibinfo  {journal} {Phys.
  Rev. Lett.}\ }\textbf {\bibinfo {volume} {100}},\ \bibinfo {pages} {175702}
  (\bibinfo {year} {2008})}\BibitemShut {NoStop}%
\bibitem [{\citenamefont {Moeckel}\ and\ \citenamefont
  {Kehrein}(2010)}]{moeckel2010}%
  \BibitemOpen
  \bibfield  {author} {\bibinfo {author} {\bibfnamefont {M.}~\bibnamefont
  {Moeckel}}\ and\ \bibinfo {author} {\bibfnamefont {S.}~\bibnamefont
  {Kehrein}},\ }\href@noop {} {\bibfield  {journal} {\bibinfo  {journal} {New
  J. Phys.}\ }\textbf {\bibinfo {volume} {12}},\ \bibinfo {pages} {055016}
  (\bibinfo {year} {2010})}\BibitemShut {NoStop}%
\bibitem [{\citenamefont {Kollar}\ \emph {et~al.}(2011)\citenamefont {Kollar},
  \citenamefont {Wolf},\ and\ \citenamefont {Eckstein}}]{kollar2011}%
  \BibitemOpen
  \bibfield  {author} {\bibinfo {author} {\bibfnamefont {M.}~\bibnamefont
  {Kollar}}, \bibinfo {author} {\bibfnamefont {F.~A.}\ \bibnamefont {Wolf}}, \
  and\ \bibinfo {author} {\bibfnamefont {M.}~\bibnamefont {Eckstein}},\
  }\href@noop {} {\bibfield  {journal} {\bibinfo  {journal} {Phys. Rev. B}\
  }\textbf {\bibinfo {volume} {84}},\ \bibinfo {pages} {054304} (\bibinfo
  {year} {2011})}\BibitemShut {NoStop}%
\bibitem [{\citenamefont {Marcuzzi}\ \emph {et~al.}(2013)\citenamefont
  {Marcuzzi}, \citenamefont {Marino}, \citenamefont {Gambassi},\ and\
  \citenamefont {Silva}}]{marcuzzi2013}%
  \BibitemOpen
  \bibfield  {author} {\bibinfo {author} {\bibfnamefont {M.}~\bibnamefont
  {Marcuzzi}}, \bibinfo {author} {\bibfnamefont {J.}~\bibnamefont {Marino}},
  \bibinfo {author} {\bibfnamefont {A.}~\bibnamefont {Gambassi}}, \ and\
  \bibinfo {author} {\bibfnamefont {A.}~\bibnamefont {Silva}},\ }\href@noop {}
  {\bibfield  {journal} {\bibinfo  {journal} {Phys. Rev. Lett.}\ }\textbf
  {\bibinfo {volume} {111}},\ \bibinfo {pages} {197203} (\bibinfo {year}
  {2013})}\BibitemShut {NoStop}%
\bibitem [{\citenamefont {Schmidt}\ and\ \citenamefont
  {Monien}()}]{schmidt2002}%
  \BibitemOpen
  \bibfield  {author} {\bibinfo {author} {\bibfnamefont {P.}~\bibnamefont
  {Schmidt}}\ and\ \bibinfo {author} {\bibfnamefont {H.}~\bibnamefont
  {Monien}},\ }\href@noop {} {\ }\Eprint
  {http://arxiv.org/abs/cond-mat/0202046} {cond-mat/0202046} \BibitemShut
  {NoStop}%
\bibitem [{\citenamefont {Freericks}\ \emph {et~al.}(2006)\citenamefont
  {Freericks}, \citenamefont {Turkowski},\ and\ \citenamefont
  {Zlati\'c}}]{freericks2006}%
  \BibitemOpen
  \bibfield  {author} {\bibinfo {author} {\bibfnamefont {J.~K.}\ \bibnamefont
  {Freericks}}, \bibinfo {author} {\bibfnamefont {V.~M.}\ \bibnamefont
  {Turkowski}}, \ and\ \bibinfo {author} {\bibfnamefont {V.}~\bibnamefont
  {Zlati\'c}},\ }\href@noop {} {\bibfield  {journal} {\bibinfo  {journal}
  {Phys. Rev. Lett.}\ }\textbf {\bibinfo {volume} {97}},\ \bibinfo {pages}
  {266408} (\bibinfo {year} {2006})}\BibitemShut {NoStop}%
\bibitem [{\citenamefont {Aoki}\ \emph {et~al.}(2014)\citenamefont {Aoki},
  \citenamefont {Tsuji}, \citenamefont {Eckstein}, \citenamefont {Kollar},
  \citenamefont {Oka},\ and\ \citenamefont {Werner}}]{aoki2013}%
  \BibitemOpen
  \bibfield  {author} {\bibinfo {author} {\bibfnamefont {H.}~\bibnamefont
  {Aoki}}, \bibinfo {author} {\bibfnamefont {N.}~\bibnamefont {Tsuji}},
  \bibinfo {author} {\bibfnamefont {M.}~\bibnamefont {Eckstein}}, \bibinfo
  {author} {\bibfnamefont {M.}~\bibnamefont {Kollar}}, \bibinfo {author}
  {\bibfnamefont {T.}~\bibnamefont {Oka}}, \ and\ \bibinfo {author}
  {\bibfnamefont {P.}~\bibnamefont {Werner}},\ }\href@noop {} {\bibfield
  {journal} {\bibinfo  {journal} {Rev. Mod. Phys.}\ }\textbf {\bibinfo {volume}
  {86}},\ \bibinfo {pages} {779} (\bibinfo {year} {2014})}\BibitemShut
  {NoStop}%
\bibitem [{\citenamefont {Eckstein}\ \emph {et~al.}(2010)\citenamefont
  {Eckstein}, \citenamefont {Kollar},\ and\ \citenamefont
  {Werner}}]{eckstein2010b}%
  \BibitemOpen
  \bibfield  {author} {\bibinfo {author} {\bibfnamefont {M.}~\bibnamefont
  {Eckstein}}, \bibinfo {author} {\bibfnamefont {M.}~\bibnamefont {Kollar}}, \
  and\ \bibinfo {author} {\bibfnamefont {P.}~\bibnamefont {Werner}},\
  }\href@noop {} {\bibfield  {journal} {\bibinfo  {journal} {Phys. Rev. B}\
  }\textbf {\bibinfo {volume} {81}},\ \bibinfo {pages} {115131} (\bibinfo
  {year} {2010})}\BibitemShut {NoStop}%
\bibitem [{\citenamefont {Hofmann}\ \emph
  {et~al.}(2016{\natexlab{a}})\citenamefont {Hofmann}, \citenamefont
  {Eckstein},\ and\ \citenamefont {Potthoff}}]{hofmann2016}%
  \BibitemOpen
  \bibfield  {author} {\bibinfo {author} {\bibfnamefont {F.}~\bibnamefont
  {Hofmann}}, \bibinfo {author} {\bibfnamefont {M.}~\bibnamefont {Eckstein}}, \
  and\ \bibinfo {author} {\bibfnamefont {M.}~\bibnamefont {Potthoff}},\
  }\href@noop {} {\bibfield  {journal} {\bibinfo  {journal} {Phys. Rev. B}\
  }\textbf {\bibinfo {volume} {93}},\ \bibinfo {pages} {235104} (\bibinfo
  {year} {2016}{\natexlab{a}})}\BibitemShut {NoStop}%
\bibitem [{\citenamefont {Sandri}\ \emph {et~al.}(2012)\citenamefont {Sandri},
  \citenamefont {Schir\'o},\ and\ \citenamefont {Fabrizio}}]{sandri2012}%
  \BibitemOpen
  \bibfield  {author} {\bibinfo {author} {\bibfnamefont {M.}~\bibnamefont
  {Sandri}}, \bibinfo {author} {\bibfnamefont {M.}~\bibnamefont {Schir\'o}}, \
  and\ \bibinfo {author} {\bibfnamefont {M.}~\bibnamefont {Fabrizio}},\
  }\href@noop {} {\bibfield  {journal} {\bibinfo  {journal} {Phys. Rev. B}\
  }\textbf {\bibinfo {volume} {86}},\ \bibinfo {pages} {075122} (\bibinfo
  {year} {2012})}\BibitemShut {NoStop}%
\bibitem [{\citenamefont {Hofmann}\ \emph {et~al.}(2013)\citenamefont
  {Hofmann}, \citenamefont {Eckstein}, \citenamefont {Arrigoni},\ and\
  \citenamefont {Potthoff}}]{hofmann2013}%
  \BibitemOpen
  \bibfield  {author} {\bibinfo {author} {\bibfnamefont {F.}~\bibnamefont
  {Hofmann}}, \bibinfo {author} {\bibfnamefont {M.}~\bibnamefont {Eckstein}},
  \bibinfo {author} {\bibfnamefont {E.}~\bibnamefont {Arrigoni}}, \ and\
  \bibinfo {author} {\bibfnamefont {M.}~\bibnamefont {Potthoff}},\ }\href@noop
  {} {\bibfield  {journal} {\bibinfo  {journal} {Phys. Rev. B}\ }\textbf
  {\bibinfo {volume} {88}},\ \bibinfo {pages} {165124} (\bibinfo {year}
  {2013})}\BibitemShut {NoStop}%
\bibitem [{\citenamefont {Potthoff}(2003{\natexlab{a}})}]{potthoff2003}%
  \BibitemOpen
  \bibfield  {author} {\bibinfo {author} {\bibfnamefont {M.}~\bibnamefont
  {Potthoff}},\ }\href@noop {} {\bibfield  {journal} {\bibinfo  {journal} {Eur.
  Phys. J. B}\ }\textbf {\bibinfo {volume} {32}},\ \bibinfo {pages} {429}
  (\bibinfo {year} {2003}{\natexlab{a}})}\BibitemShut {NoStop}%
\bibitem [{\citenamefont {Potthoff}\ \emph {et~al.}(2003)\citenamefont
  {Potthoff}, \citenamefont {Aichhorn},\ and\ \citenamefont
  {Dahnken}}]{potthoff2003c}%
  \BibitemOpen
  \bibfield  {author} {\bibinfo {author} {\bibfnamefont {M.}~\bibnamefont
  {Potthoff}}, \bibinfo {author} {\bibfnamefont {M.}~\bibnamefont {Aichhorn}},
  \ and\ \bibinfo {author} {\bibfnamefont {C.}~\bibnamefont {Dahnken}},\
  }\href@noop {} {\bibfield  {journal} {\bibinfo  {journal} {Phys. Rev. Lett.}\
  }\textbf {\bibinfo {volume} {91}},\ \bibinfo {pages} {206402} (\bibinfo
  {year} {2003})}\BibitemShut {NoStop}%
\bibitem [{\citenamefont {Werner}\ and\ \citenamefont
  {Eckstein}(2012)}]{werner2012}%
  \BibitemOpen
  \bibfield  {author} {\bibinfo {author} {\bibfnamefont {P.}~\bibnamefont
  {Werner}}\ and\ \bibinfo {author} {\bibfnamefont {M.}~\bibnamefont
  {Eckstein}},\ }\href@noop {} {\bibfield  {journal} {\bibinfo  {journal}
  {Phys. Rev. B}\ }\textbf {\bibinfo {volume} {86}},\ \bibinfo {pages} {045119}
  (\bibinfo {year} {2012})}\BibitemShut {NoStop}%
\bibitem [{\citenamefont {Werner}\ and\ \citenamefont
  {Eckstein}(2015)}]{werner2015}%
  \BibitemOpen
  \bibfield  {author} {\bibinfo {author} {\bibfnamefont {P.}~\bibnamefont
  {Werner}}\ and\ \bibinfo {author} {\bibfnamefont {M.}~\bibnamefont
  {Eckstein}},\ }\href@noop {} {\bibfield  {journal} {\bibinfo  {journal}
  {Europhys. Lett.}\ }\textbf {\bibinfo {volume} {109}},\ \bibinfo {pages}
  {37002} (\bibinfo {year} {2015})}\BibitemShut {NoStop}%
\bibitem [{\citenamefont {Gull}\ \emph {et~al.}(2011)\citenamefont {Gull},
  \citenamefont {Millis}, \citenamefont {Lichtenstein}, \citenamefont
  {Rubtsov}, \citenamefont {Troyer},\ and\ \citenamefont {Werner}}]{gull2011}%
  \BibitemOpen
  \bibfield  {author} {\bibinfo {author} {\bibfnamefont {E.}~\bibnamefont
  {Gull}}, \bibinfo {author} {\bibfnamefont {A.~J.}\ \bibnamefont {Millis}},
  \bibinfo {author} {\bibfnamefont {A.~I.}\ \bibnamefont {Lichtenstein}},
  \bibinfo {author} {\bibfnamefont {A.~N.}\ \bibnamefont {Rubtsov}}, \bibinfo
  {author} {\bibfnamefont {M.}~\bibnamefont {Troyer}}, \ and\ \bibinfo {author}
  {\bibfnamefont {P.}~\bibnamefont {Werner}},\ }\href@noop {} {\bibfield
  {journal} {\bibinfo  {journal} {Rev. Mod. Phys.}\ }\textbf {\bibinfo {volume}
  {83}},\ \bibinfo {pages} {349} (\bibinfo {year} {2011})}\BibitemShut
  {NoStop}%
\bibitem [{\citenamefont {Gramsch}\ \emph {et~al.}(2013)\citenamefont
  {Gramsch}, \citenamefont {Balzer}, \citenamefont {Eckstein},\ and\
  \citenamefont {Kollar}}]{gramsch2013}%
  \BibitemOpen
  \bibfield  {author} {\bibinfo {author} {\bibfnamefont {C.}~\bibnamefont
  {Gramsch}}, \bibinfo {author} {\bibfnamefont {K.}~\bibnamefont {Balzer}},
  \bibinfo {author} {\bibfnamefont {M.}~\bibnamefont {Eckstein}}, \ and\
  \bibinfo {author} {\bibfnamefont {M.}~\bibnamefont {Kollar}},\ }\href@noop {}
  {\bibfield  {journal} {\bibinfo  {journal} {Phys. Rev. B}\ }\textbf {\bibinfo
  {volume} {88}},\ \bibinfo {pages} {235106} (\bibinfo {year}
  {2013})}\BibitemShut {NoStop}%
\bibitem [{\citenamefont {Balzer}\ \emph {et~al.}(2015)\citenamefont {Balzer},
  \citenamefont {Li}, \citenamefont {Vendrell},\ and\ \citenamefont
  {Eckstein}}]{balzer2015}%
  \BibitemOpen
  \bibfield  {author} {\bibinfo {author} {\bibfnamefont {K.}~\bibnamefont
  {Balzer}}, \bibinfo {author} {\bibfnamefont {Z.}~\bibnamefont {Li}}, \bibinfo
  {author} {\bibfnamefont {O.}~\bibnamefont {Vendrell}}, \ and\ \bibinfo
  {author} {\bibfnamefont {M.}~\bibnamefont {Eckstein}},\ }\href@noop {}
  {\bibfield  {journal} {\bibinfo  {journal} {Phys. Rev. B}\ }\textbf {\bibinfo
  {volume} {91}},\ \bibinfo {pages} {045136} (\bibinfo {year}
  {2015})}\BibitemShut {NoStop}%
\bibitem [{\citenamefont {Wolf}\ \emph {et~al.}(2014)\citenamefont {Wolf},
  \citenamefont {McCulloch},\ and\ \citenamefont {Schollw\"ock}}]{wolf2014}%
  \BibitemOpen
  \bibfield  {author} {\bibinfo {author} {\bibfnamefont {F.~A.}\ \bibnamefont
  {Wolf}}, \bibinfo {author} {\bibfnamefont {I.~P.}\ \bibnamefont {McCulloch}},
  \ and\ \bibinfo {author} {\bibfnamefont {U.}~\bibnamefont {Schollw\"ock}},\
  }\href@noop {} {\bibfield  {journal} {\bibinfo  {journal} {Phys. Rev. B}\
  }\textbf {\bibinfo {volume} {90}},\ \bibinfo {pages} {235131} (\bibinfo
  {year} {2014})}\BibitemShut {NoStop}%
\bibitem [{\citenamefont {Balzer}\ and\ \citenamefont
  {Potthoff}(2011)}]{potthoff2011c}%
  \BibitemOpen
  \bibfield  {author} {\bibinfo {author} {\bibfnamefont {M.}~\bibnamefont
  {Balzer}}\ and\ \bibinfo {author} {\bibfnamefont {M.}~\bibnamefont
  {Potthoff}},\ }\href@noop {} {\bibfield  {journal} {\bibinfo  {journal}
  {Phys. Rev. B}\ }\textbf {\bibinfo {volume} {83}},\ \bibinfo {pages} {195132}
  (\bibinfo {year} {2011})}\BibitemShut {NoStop}%
\bibitem [{\citenamefont {Gramsch}\ and\ \citenamefont
  {Potthoff}(2015)}]{gramsch2015}%
  \BibitemOpen
  \bibfield  {author} {\bibinfo {author} {\bibfnamefont {C.}~\bibnamefont
  {Gramsch}}\ and\ \bibinfo {author} {\bibfnamefont {M.}~\bibnamefont
  {Potthoff}},\ }\href@noop {} {\bibfield  {journal} {\bibinfo  {journal}
  {Phys. Rev. B}\ }\textbf {\bibinfo {volume} {92}},\ \bibinfo {pages} {235135}
  (\bibinfo {year} {2015})}\BibitemShut {NoStop}%
\bibitem [{\citenamefont {Behrmann}\ \emph {et~al.}(2013)\citenamefont
  {Behrmann}, \citenamefont {Fabrizio},\ and\ \citenamefont
  {Lechermann}}]{behrmann2013}%
  \BibitemOpen
  \bibfield  {author} {\bibinfo {author} {\bibfnamefont {M.}~\bibnamefont
  {Behrmann}}, \bibinfo {author} {\bibfnamefont {M.}~\bibnamefont {Fabrizio}},
  \ and\ \bibinfo {author} {\bibfnamefont {F.}~\bibnamefont {Lechermann}},\
  }\href@noop {} {\bibfield  {journal} {\bibinfo  {journal} {Phys. Rev. B}\
  }\textbf {\bibinfo {volume} {88}},\ \bibinfo {pages} {035116} (\bibinfo
  {year} {2013})}\BibitemShut {NoStop}%
\bibitem [{\citenamefont {Huscroft}\ \emph {et~al.}(1999)\citenamefont
  {Huscroft}, \citenamefont {McMahan},\ and\ \citenamefont
  {Scalettar}}]{huscroft1999}%
  \BibitemOpen
  \bibfield  {author} {\bibinfo {author} {\bibfnamefont {C.}~\bibnamefont
  {Huscroft}}, \bibinfo {author} {\bibfnamefont {A.~K.}\ \bibnamefont
  {McMahan}}, \ and\ \bibinfo {author} {\bibfnamefont {R.~T.}\ \bibnamefont
  {Scalettar}},\ }\href@noop {} {\bibfield  {journal} {\bibinfo  {journal}
  {Phys. Rev. Lett.}\ }\textbf {\bibinfo {volume} {82}},\ \bibinfo {pages}
  {2342} (\bibinfo {year} {1999})}\BibitemShut {NoStop}%
\bibitem [{\citenamefont {Held}\ \emph {et~al.}(2000)\citenamefont {Held},
  \citenamefont {Huscroft}, \citenamefont {Scalettar},\ and\ \citenamefont
  {McMahan}}]{held2000}%
  \BibitemOpen
  \bibfield  {author} {\bibinfo {author} {\bibfnamefont {K.}~\bibnamefont
  {Held}}, \bibinfo {author} {\bibfnamefont {C.}~\bibnamefont {Huscroft}},
  \bibinfo {author} {\bibfnamefont {R.~T.}\ \bibnamefont {Scalettar}}, \ and\
  \bibinfo {author} {\bibfnamefont {A.~K.}\ \bibnamefont {McMahan}},\
  }\href@noop {} {\bibfield  {journal} {\bibinfo  {journal} {Phys. Rev. Lett.}\
  }\textbf {\bibinfo {volume} {85}},\ \bibinfo {pages} {373} (\bibinfo {year}
  {2000})}\BibitemShut {NoStop}%
\bibitem [{\citenamefont {Held}\ and\ \citenamefont {Bulla}(2000)}]{held2000b}%
  \BibitemOpen
  \bibfield  {author} {\bibinfo {author} {\bibfnamefont {K.}~\bibnamefont
  {Held}}\ and\ \bibinfo {author} {\bibfnamefont {R.}~\bibnamefont {Bulla}},\
  }\href@noop {} {\bibfield  {journal} {\bibinfo  {journal} {Euro. Phys. J. B}\
  }\textbf {\bibinfo {volume} {17}},\ \bibinfo {pages} {7} (\bibinfo {year}
  {2000})}\BibitemShut {NoStop}%
\bibitem [{\citenamefont {Dongen}\ \emph {et~al.}(2001)\citenamefont {Dongen},
  \citenamefont {Majumdar}, \citenamefont {Huscroft},\ and\ \citenamefont
  {Zhang}}]{vandongen2001}%
  \BibitemOpen
  \bibfield  {author} {\bibinfo {author} {\bibfnamefont {P.~v.}\ \bibnamefont
  {Dongen}}, \bibinfo {author} {\bibfnamefont {K.}~\bibnamefont {Majumdar}},
  \bibinfo {author} {\bibfnamefont {C.}~\bibnamefont {Huscroft}}, \ and\
  \bibinfo {author} {\bibfnamefont {F.-C.}\ \bibnamefont {Zhang}},\ }\href@noop
  {} {\bibfield  {journal} {\bibinfo  {journal} {Phys. Rev. B}\ }\textbf
  {\bibinfo {volume} {64}},\ \bibinfo {pages} {195123} (\bibinfo {year}
  {2001})}\BibitemShut {NoStop}%
\bibitem [{\citenamefont {Bulla}\ and\ \citenamefont
  {Potthoff}(2000)}]{bulla2000}%
  \BibitemOpen
  \bibfield  {author} {\bibinfo {author} {\bibfnamefont {R.}~\bibnamefont
  {Bulla}}\ and\ \bibinfo {author} {\bibfnamefont {M.}~\bibnamefont
  {Potthoff}},\ }\href@noop {} {\bibfield  {journal} {\bibinfo  {journal}
  {Euro. Phys. J. B}\ }\textbf {\bibinfo {volume} {13}},\ \bibinfo {pages}
  {257} (\bibinfo {year} {2000})}\BibitemShut {NoStop}%
\bibitem [{\citenamefont {Brinkman}\ and\ \citenamefont
  {Rice}(1970)}]{brinkman1970}%
  \BibitemOpen
  \bibfield  {author} {\bibinfo {author} {\bibfnamefont {W.~F.}\ \bibnamefont
  {Brinkman}}\ and\ \bibinfo {author} {\bibfnamefont {T.~M.}\ \bibnamefont
  {Rice}},\ }\href@noop {} {\bibfield  {journal} {\bibinfo  {journal} {Phys.
  Rev. B}\ }\textbf {\bibinfo {volume} {2}},\ \bibinfo {pages} {4302} (\bibinfo
  {year} {1970})}\BibitemShut {NoStop}%
\bibitem [{\citenamefont {Hofmann}\ \emph
  {et~al.}(2016{\natexlab{b}})\citenamefont {Hofmann}, \citenamefont
  {Eckstein},\ and\ \citenamefont {Potthoff}}]{hofmann2015}%
  \BibitemOpen
  \bibfield  {author} {\bibinfo {author} {\bibfnamefont {F.}~\bibnamefont
  {Hofmann}}, \bibinfo {author} {\bibfnamefont {M.}~\bibnamefont {Eckstein}}, \
  and\ \bibinfo {author} {\bibfnamefont {M.}~\bibnamefont {Potthoff}},\
  }\href@noop {} {\bibfield  {journal} {\bibinfo  {journal} {J. Phys.: Conf.
  Series}\ }\textbf {\bibinfo {volume} {696}},\ \bibinfo {pages} {012002}
  (\bibinfo {year} {2016}{\natexlab{b}})}\BibitemShut {NoStop}%
\bibitem [{\citenamefont {van Leeuwen}\ \emph {et~al.}(2006)\citenamefont {van
  Leeuwen}, \citenamefont {Dahlen}, \citenamefont {Stefanucci}, \citenamefont
  {Almbladh},\ and\ \citenamefont {von Barth}}]{leeuwen2006c}%
  \BibitemOpen
  \bibfield  {author} {\bibinfo {author} {\bibfnamefont {R.}~\bibnamefont {van
  Leeuwen}}, \bibinfo {author} {\bibfnamefont {N.~E.}\ \bibnamefont {Dahlen}},
  \bibinfo {author} {\bibfnamefont {G.}~\bibnamefont {Stefanucci}}, \bibinfo
  {author} {\bibfnamefont {C.~O.}\ \bibnamefont {Almbladh}}, \ and\ \bibinfo
  {author} {\bibfnamefont {U.}~\bibnamefont {von Barth}},\ }in\ \href@noop {}
  {\emph {\bibinfo {booktitle} {Time-Dependent Density Functional Theory}}},\
  \bibinfo {editor} {edited by\ \bibinfo {editor} {\bibfnamefont {M.~A.~L.}\
  \bibnamefont {Marques}}, \bibinfo {editor} {\bibfnamefont {C.~A.}\
  \bibnamefont {Ullrich}}, \bibinfo {editor} {\bibfnamefont {F.}~\bibnamefont
  {Nogueira}}, \bibinfo {editor} {\bibfnamefont {A.}~\bibnamefont {Rubio}},
  \bibinfo {editor} {\bibfnamefont {K.}~\bibnamefont {Burke}}, \ and\ \bibinfo
  {editor} {\bibfnamefont {E.~K.~U.}\ \bibnamefont {Gross}}}\ (\bibinfo
  {publisher} {Springer},\ \bibinfo {address} {Berlin Heidelberg},\ \bibinfo
  {year} {2006})\ p.~\bibinfo {pages} {33}\BibitemShut {NoStop}%
\bibitem [{\citenamefont {Rammer}(2007)}]{rammer2007}%
  \BibitemOpen
  \bibfield  {author} {\bibinfo {author} {\bibfnamefont {J.}~\bibnamefont
  {Rammer}},\ }\href@noop {} {\emph {\bibinfo {title} {Quantum field theory of
  nonequilibrium states}}}\ (\bibinfo  {publisher} {Cambridge University
  Press},\ \bibinfo {year} {2007})\BibitemShut {NoStop}%
\bibitem [{\citenamefont {Kelley}(1987)}]{kelley1987}%
  \BibitemOpen
  \bibfield  {author} {\bibinfo {author} {\bibfnamefont {C.~T.}\ \bibnamefont
  {Kelley}},\ }\href@noop {} {\emph {\bibinfo {title} {Solving nonlinear
  equations with Newton's method}}},\ Fundamentals of algorithms\ (\bibinfo
  {publisher} {Society for Industrial and Applied Mathematics},\ \bibinfo
  {year} {1987})\BibitemShut {NoStop}%
\bibitem [{\citenamefont {Broyden}(1965)}]{broyden1965}%
  \BibitemOpen
  \bibfield  {author} {\bibinfo {author} {\bibfnamefont {C.~G.}\ \bibnamefont
  {Broyden}},\ }\href@noop {} {\bibfield  {journal} {\bibinfo  {journal} {Math.
  Comp.}\ }\textbf {\bibinfo {volume} {19}},\ \bibinfo {pages} {577} (\bibinfo
  {year} {1965})}\BibitemShut {NoStop}%
\bibitem [{\citenamefont {Lange}(1998)}]{lange1998}%
  \BibitemOpen
  \bibfield  {author} {\bibinfo {author} {\bibfnamefont {E.}~\bibnamefont
  {Lange}},\ }\href@noop {} {\bibfield  {journal} {\bibinfo  {journal} {Modern
  Physics Letters B}\ }\textbf {\bibinfo {volume} {12}},\ \bibinfo {pages}
  {915} (\bibinfo {year} {1998})}\BibitemShut {NoStop}%
\bibitem [{\citenamefont {Potthoff}(2003{\natexlab{b}})}]{potthoff2003b}%
  \BibitemOpen
  \bibfield  {author} {\bibinfo {author} {\bibfnamefont {M.}~\bibnamefont
  {Potthoff}},\ }\href@noop {} {\bibfield  {journal} {\bibinfo  {journal} {Eur.
  Phys. J. B}\ }\textbf {\bibinfo {volume} {36}},\ \bibinfo {pages} {335}
  (\bibinfo {year} {2003}{\natexlab{b}})}\BibitemShut {NoStop}%
\bibitem [{\citenamefont {Georges}\ and\ \citenamefont
  {Krauth}(1993)}]{georges1993}%
  \BibitemOpen
  \bibfield  {author} {\bibinfo {author} {\bibfnamefont {A.}~\bibnamefont
  {Georges}}\ and\ \bibinfo {author} {\bibfnamefont {W.}~\bibnamefont
  {Krauth}},\ }\href@noop {} {\bibfield  {journal} {\bibinfo  {journal} {Phys.
  Rev. B}\ }\textbf {\bibinfo {volume} {48}},\ \bibinfo {pages} {7167}
  (\bibinfo {year} {1993})}\BibitemShut {NoStop}%
\bibitem [{\citenamefont {Rozenberg}\ \emph {et~al.}(1994)\citenamefont
  {Rozenberg}, \citenamefont {Kotliar},\ and\ \citenamefont
  {Zhang}}]{rozenberg1994}%
  \BibitemOpen
  \bibfield  {author} {\bibinfo {author} {\bibfnamefont {M.~J.}\ \bibnamefont
  {Rozenberg}}, \bibinfo {author} {\bibfnamefont {G.}~\bibnamefont {Kotliar}},
  \ and\ \bibinfo {author} {\bibfnamefont {X.~Y.}\ \bibnamefont {Zhang}},\
  }\href@noop {} {\bibfield  {journal} {\bibinfo  {journal} {Phys. Rev. B}\
  }\textbf {\bibinfo {volume} {49}},\ \bibinfo {pages} {10181} (\bibinfo {year}
  {1994})}\BibitemShut {NoStop}%
\bibitem [{\citenamefont {Po\v{z}gaj\v{c}i\'{c}}(2004)}]{pozgajcic2004}%
  \BibitemOpen
  \bibfield  {author} {\bibinfo {author} {\bibfnamefont {K.}~\bibnamefont
  {Po\v{z}gaj\v{c}i\'{c}}},\ }\href@noop {} {cond-mat/0407172} \BibitemShut
  {NoStop}%
\bibitem [{\citenamefont {Strohmaier}\ \emph {et~al.}(2010)\citenamefont
  {Strohmaier}, \citenamefont {Greif}, \citenamefont {J\"ordens}, \citenamefont
  {Tarruell}, \citenamefont {Moritz}, \citenamefont {Esslinger}, \citenamefont
  {Sensarma}, \citenamefont {Pekker}, \citenamefont {Altman},\ and\
  \citenamefont {Demler}}]{strohmaier2010}%
  \BibitemOpen
  \bibfield  {author} {\bibinfo {author} {\bibfnamefont {N.}~\bibnamefont
  {Strohmaier}}, \bibinfo {author} {\bibfnamefont {D.}~\bibnamefont {Greif}},
  \bibinfo {author} {\bibfnamefont {R.}~\bibnamefont {J\"ordens}}, \bibinfo
  {author} {\bibfnamefont {L.}~\bibnamefont {Tarruell}}, \bibinfo {author}
  {\bibfnamefont {H.}~\bibnamefont {Moritz}}, \bibinfo {author} {\bibfnamefont
  {T.}~\bibnamefont {Esslinger}}, \bibinfo {author} {\bibfnamefont
  {R.}~\bibnamefont {Sensarma}}, \bibinfo {author} {\bibfnamefont
  {D.}~\bibnamefont {Pekker}}, \bibinfo {author} {\bibfnamefont
  {E.}~\bibnamefont {Altman}}, \ and\ \bibinfo {author} {\bibfnamefont
  {E.}~\bibnamefont {Demler}},\ }\href@noop {} {\bibfield  {journal} {\bibinfo
  {journal} {Phys. Rev. Lett.}\ }\textbf {\bibinfo {volume} {104}},\ \bibinfo
  {pages} {080401} (\bibinfo {year} {2010})}\BibitemShut {NoStop}%
\bibitem [{\citenamefont {Hofmann}\ and\ \citenamefont
  {Potthoff}(2012)}]{hofmann2012}%
  \BibitemOpen
  \bibfield  {author} {\bibinfo {author} {\bibfnamefont {F.}~\bibnamefont
  {Hofmann}}\ and\ \bibinfo {author} {\bibfnamefont {M.}~\bibnamefont
  {Potthoff}},\ }\href@noop {} {\bibfield  {journal} {\bibinfo  {journal}
  {Phys. Rev. B}\ }\textbf {\bibinfo {volume} {85}},\ \bibinfo {pages} {205127}
  (\bibinfo {year} {2012})}\BibitemShut {NoStop}%
\bibitem [{\citenamefont {Rausch}\ and\ \citenamefont {Potthoff}(2016)}]{RP16}%
  \BibitemOpen
  \bibfield  {author} {\bibinfo {author} {\bibfnamefont {R.}~\bibnamefont
  {Rausch}}\ and\ \bibinfo {author} {\bibfnamefont {M.}~\bibnamefont
  {Potthoff}},\ }\href@noop {} {\bibfield  {journal} {\bibinfo  {journal} {New
  J. Phys}\ }\textbf {\bibinfo {volume} {18}},\ \bibinfo {pages} {023033}
  (\bibinfo {year} {2016})}\BibitemShut {NoStop}%
\bibitem [{\citenamefont {Schir\'o}\ and\ \citenamefont
  {Fabrizio}(2011)}]{schiro2011}%
  \BibitemOpen
  \bibfield  {author} {\bibinfo {author} {\bibfnamefont {M.}~\bibnamefont
  {Schir\'o}}\ and\ \bibinfo {author} {\bibfnamefont {M.}~\bibnamefont
  {Fabrizio}},\ }\href@noop {} {\bibfield  {journal} {\bibinfo  {journal}
  {Phys. Rev. B}\ }\textbf {\bibinfo {volume} {83}},\ \bibinfo {pages} {165105}
  (\bibinfo {year} {2011})}\BibitemShut {NoStop}%
\bibitem [{\citenamefont {Dziarmaga}(2010)}]{dziarmaga2010}%
  \BibitemOpen
  \bibfield  {author} {\bibinfo {author} {\bibfnamefont {J.}~\bibnamefont
  {Dziarmaga}},\ }\href@noop {} {\bibfield  {journal} {\bibinfo  {journal}
  {Adv. Phys.}\ }\textbf {\bibinfo {volume} {59}},\ \bibinfo {pages} {1063}
  (\bibinfo {year} {2010})}\BibitemShut {NoStop}%
\bibitem [{\citenamefont {Francuz}\ \emph {et~al.}(2016)\citenamefont
  {Francuz}, \citenamefont {Dziarmaga}, \citenamefont {Gardas},\ and\
  \citenamefont {Zurek}}]{dziarmaga2016}%
  \BibitemOpen
  \bibfield  {author} {\bibinfo {author} {\bibfnamefont {A.}~\bibnamefont
  {Francuz}}, \bibinfo {author} {\bibfnamefont {J.}~\bibnamefont {Dziarmaga}},
  \bibinfo {author} {\bibfnamefont {B.}~\bibnamefont {Gardas}}, \ and\ \bibinfo
  {author} {\bibfnamefont {W.~H.}\ \bibnamefont {Zurek}},\ }\href@noop {}
  {\bibfield  {journal} {\bibinfo  {journal} {Phys. Rev. B}\ }\textbf {\bibinfo
  {volume} {93}},\ \bibinfo {pages} {075134} (\bibinfo {year}
  {2016})}\BibitemShut {NoStop}%
\bibitem [{\citenamefont {Moeller}\ \emph {et~al.}(1995)\citenamefont
  {Moeller}, \citenamefont {Si}, \citenamefont {Kotliar}, \citenamefont
  {Rozenberg},\ and\ \citenamefont {Fisher}}]{Moll95}%
  \BibitemOpen
  \bibfield  {author} {\bibinfo {author} {\bibfnamefont {G.}~\bibnamefont
  {Moeller}}, \bibinfo {author} {\bibfnamefont {Q.}~\bibnamefont {Si}},
  \bibinfo {author} {\bibfnamefont {G.}~\bibnamefont {Kotliar}}, \bibinfo
  {author} {\bibfnamefont {M.}~\bibnamefont {Rozenberg}}, \ and\ \bibinfo
  {author} {\bibfnamefont {D.~S.}\ \bibnamefont {Fisher}},\ }\href@noop {}
  {\bibfield  {journal} {\bibinfo  {journal} {Phys. Rev. Lett.}\ }\textbf
  {\bibinfo {volume} {74}},\ \bibinfo {pages} {2082} (\bibinfo {year}
  {1995})}\BibitemShut {NoStop}%
\end{thebibliography}

%

\end{document}